\documentclass[12pt]{article}
\usepackage{cite,graphicx,epsfig}
\newcommand{\ba}{\begin{array}}
\newcommand{\ea}{\end{array}}
\newcommand{\be}{\begin{equation}}
\newcommand{\ee}{\end{equation}}
\newcommand{\bea}{\begin{eqnarray}}
\newcommand{\eea}{\end{eqnarray}}

\def\IB{\relax\hbox{$\inbar\kern-.3em{\rm B}$}}
\def\IC{\relax\hbox{$\inbar\kern-.3em{\rm C}$}}
\def\ID{\relax\hbox{$\inbar\kern-.3em{\rm D}$}}
\def\IE{\relax\hbox{$\inbar\kern-.3em{\rm E}$}}
\def\IF{\relax\hbox{$\inbar\kern-.3em{\rm F}$}}
\def\IG{\relax\hbox{$\inbar\kern-.3em{\rm G}$}}
\def\IGa{\relax\hbox{${\rm I}\kern-.18em\Gamma$}}
\def\IH{\relax{\rm I\kern-.18em H}}
\def\IK{\relax{\rm I\kern-.18em K}}
\def\IL{\relax{\rm I\kern-.18em L}}
\def\IP{\relax{\rm I\kern-.18em P}}
\def\IR{\relax{\rm I\kern-.18em R}}
\def\IZ{\relax{\rm Z\kern-.5em Z}}

\def\half{\frac{1}{2}}

\def\f{\frac}

\def\TL{Temperley-Lieb }

\begin{document}
\begin{titlepage}

\begin{flushright}

\end{flushright}

\vskip 2 cm

\begin{center}
{\LARGE The \TL algebra and its generalizations in the Potts and XXZ models}
\vskip 1 cm

{\large A. Nichols\footnote{nichols@th.physik.uni-bonn.de} }

\begin{center}
{\em $^{1,2}$Physikalisches Institut der Universit\"at Bonn, \\
 Nussallee 12, 53115 Bonn, Germany.}
\vskip 1 cm

\end{center}

\vskip 1 cm

\begin{abstract}
We discuss generalizations of the \TL algebra in the Potts and XXZ
models. These can be used to describe the addition of different types of integrable boundary terms.

We use the \TL algebra and its one-boundary, two-boundary, and periodic
extensions to classify different integrable boundary terms in the $2$, $3$,
and $4$-state Potts models. The representations always lie at critical points where the
algebras becomes non-semisimple and possess indecomposable
representations. In
the one-boundary case we show how to use representation theory to extract the
Potts spectrum from an XXZ model with particular boundary terms and hence
obtain the finite size scaling of the Potts models. In the two-boundary case
we find that the Potts spectrum can be obtained by combining several XXZ
models with different boundary terms. As in the \TL case there is a direct
correspondence between 
representations of the lattice algebra and those in the
continuum conformal field theory.

\end{abstract}

\end{center}

\end{titlepage}

\section{Introduction}
\renewcommand{\theequation}{\arabic{section}.\arabic{equation}}
\setcounter{equation}{0}  
In this paper we shall use generalizations of the \TL algebra to discuss the XXZ and Potts models with different types of integrable boundary terms. This will allow us to find relations between the spectra of these models and hence derive finite size scaling results in the Potts models from known analytical results in the XXZ model.

The Ising model is probably the best
studied statistical model in two dimensions. The $Q$-state Potts models are a generalization of this in which the symmetry group is the symmetric
group, $S_Q$. We shall study the cases $Q=2,3,4$ which are known to have a second
order phase transition \cite{BaxterBook}. The XXZ model is another extremely
well-studied integrable model. In this paper we shall discuss representations of the \TL (TL) algebra
\cite{TemperleyLieb,MartinBook} and its one-boundary, two-boundary, and
periodic extensions \cite{JanReview} which appear in the XXZ and Potts models. These are of
interest as both the XXZ and Potts models can be written in terms of the
\emph{same} algebraic Hamiltonian. The XXZ model can be solved using Bethe
Ansatz techniques \cite{Alcaraz:1987ix,Alcaraz:1988zr,Alcaraz:1987uk}. As we
shall see in many cases the XXZ spectra contains all the Potts spectra and using the underlying algebra the spectra and finite size
scaling behaviour of the Potts models can now be extracted, in a completely
controlled way, from that of the XXZ models. 

In section \ref{sec:TL} we review the generators of the TL algebra in
the XXZ and Potts representations. The TL algebra depends on a single
parameter $q$. In the $2$, $3$, and $4$-state Potts representations
$q$ takes the values $e^{\pi i/4}$, $e^{\pi i/6}$, and $1$
respectively. In contrast the XXZ representation exists for any value of
$q$. The \TL algebra is semisimple for generic values of $q$ and all representations are fully decomposable. However for exceptional points when $q$ is a root of unity (relevant for Potts models) the algebra becomes non-semisimple and possesses both `good' irreducible and `bad' indecomposable
representations. For the XXZ representation this structure can also be
understood by studying the centralizer, the quantum group
$SU_q(2)$, which commutes with each generator\cite{Pasquier:1989kd}.

The $L$-site $2$, $3$, and $4$-state Potts models with free
boundary conditions and the $2L$-site XXZ
model with $SU_q(2)$ boundary conditions at the points $q=e^{\pi i/4}$ and
$q=e^{\pi i/6}$ and $q=1$ can be written in terms of \emph{one} Hamiltonian in two different representations of the TL
algebra. The XXZ representation of TL is known to be faithful
\cite{Martin:1991zk} and therefore
evaluating a TL Hamiltonian in the XXZ representation gives \emph{all} the
eigenvalues allowed by the algebra and any other representation (e.g. Potts)
must give, up to degeneracies, some subset of the XXZ eigenvalues. This explains the numerical coincidences found in \cite{Alcaraz:1987ix,Alcaraz:1988zr,Alcaraz:1987uk}. The fact that these spectral coincidences continue to be found for
inhomogeneous chains emphasizes their algebraic, rather than integrable,
nature. It is only the energy levels of the `good' states of the  $SU_q(2)$
invariant XXZ models that are present in the Potts models with free boundary conditions.

An interesting generalization of the TL algebra known as the one-boundary TL
(1BTL), or blob, algebra has been well studied in the mathematical literature
\cite{Martin:1992td,Martin:1993jk,MartinWoodcockI,MartinWoodcockII}. The 1BTL
algebra depends on two parameters, one for the bulk generators and one for the
boundary generator. In section \ref{sec:1BTLrepns} we give the Potts and XXZ
representations of the 1BTL algebra. The XXZ representation involves a non-diagonal boundary operator. It exists for general
values of the parameters and its structure can also be understood by a
detailed study \cite{Nichols:2004fb} of its centralizer\cite{Delius:2001qh,Doikou:2004km}. In the Potts representations we shall see how the bulk symmetry can be broken in different ways by the presence of integrable boundary terms.

In section \ref{se:1BTLrepntheory} we discuss the representation theory of the
1BTL algebra\cite{Martin:1992td,Martin:1993jk,MartinWoodcockI,MartinWoodcockII}. As in the quantum group case the values of the parameters
realized in the Potts representations correspond to exceptional cases of the
1BTL algebra in which one has both `good' irreducible and `bad' indecomposable
representations. In section \ref{sec:1BTLinPotts} we show that the energy
levels of the $L$-site $2$ and $3$-state Potts models with different types of
integrable 
boundary terms added to one end can be found within the $2L$-site XXZ
models with a single non-diagonal boundary term. The explanation of this phenomena is similar to the TL case - here they are simply
two different representations of the 1BTL algebra. The
energy levels from the XXZ chain which occur in the Potts models 
correspond to the `good' representations of the 1BTL algebra. In contrast to the TL case the issue of faithfulness of the XXZ
representation is less clear \cite{PaulMartinPrivateCommunication}. The results of this paper concern
only the irreducible representations. They are compatible with the
statement that the XXZ representation contains all possible irreducible
structure allowed by the 1BTL algebra. In \cite{Nichols:2004fb} we found a spectral
equivalence between the XXZ model with arbitrary boundary term added to one
end and the same XXZ model with diagonal boundary terms. Therefore an
understanding of the one-boundary chain immediately allows one to understand
all of the numerical results for obtaining Potts spectra from diagonal chains \cite{Alcaraz:1987uk,Alcaraz:1988ey}. 

In section \ref{sec:2BTL} we discuss a further generalization of the TL
algebra: the two-boundary (2BTL) algebra. This algebra is different
from the TL and 1BTL algebras in that the number of
words is no longer finite and the representation theory is essentially unknown. All representations
that occur in this paper lie in a finite dimensional quotient that involves an additional parameter
$b$. In \cite{deGier:2005fx} the 2BTL algebra with this quotient was studied
and a conjecture was made for the values of the parameter $b$ that correspond
to exceptional points of the 2BTL algebra. The Potts
representations correspond to very particular values of the parameters all of which are exceptional. The XXZ representation exists
for arbitrary values of $b$ and the parameters in the algebra. However, in
contrast to the TL and 1BTL cases, it is certainly not faithful (as we shall
see explicitly at two sites) and so we cannot hope to obtain all the
Potts eigenvalues from a single XXZ chain. However we shall see that by \emph{combining} sectors of XXZ chains with different exceptional
values of $b$ one can get all of the Potts eigenvalues. Here we shall give numerical results and sketch the mechanism
that is responsible. A systematic procedure for extracting them requires a
knowledge of 2BTL representation theory and we shall not describe this here. In certain cases however the
2BTL representations lie in a quotient of the TL, or 1BTL, algebra and one can understand the representations and hence extract
the Potts spectra from XXZ ones. This generalizes the results of
\cite{Alcaraz:2000hv,BelavinUsmanov} for the Ising case and allows one to expand and cross-check many results. 

In section \ref{se:FSS} we shall discuss the finite size scaling (FSS) of the
one-boundary chains. A central result of \cite{Nichols:2004fb} concerning
the spectral equivalence of the one-boundary and diagonal chains will be of
great practical value as many results are known about the FSS limit of the
diagonal chain\cite{deVega:1984qb,Hamer,HamerBatchelor,Hamer:1987ei}. All results can be understood algebraically in terms of the
finite size scaling of an integrable 1BTL Hamiltonian acting on a
particular (irreducible) representation. The structure of irreducible representations of the 1BTL is related, in the
finite size scaling limit, to the embedding of Verma modules in the
Virasoro minimal models\cite{Rocha-Caridi}. One benefit of this algebraic point of view is that it does not depend on the use of a particular representation. We shall use the Potts representations of 1BTL to derive FSS results for the Ising and $3$-state Potts
models with boundary terms. Our results are in complete agreement with previous studies
\cite{Alcaraz:1987uk,Alcaraz:1988zr} and continuum results \cite{Cardy:1989ir,Affleck:1998nq}. In \cite{Koo:1992pe} generalizations of the Potts models were constructed which have extended multi-site interactions. The finite size scaling of these theories were also discussed from the point of view of the TL algebra.

In section \ref{sec:PTL} we discuss briefly the periodic Temperley-Lieb (PTL)
algebra\cite{Levy:1991nc,Martin:1992td,Martin:1993jk}. This algebra has also
attracted attention in the mathematical literature
\cite{GrahamLehrer,JonesReznikoff}. The algebra again has an
infinite number of words however, as with the
2BTL case, all representations occurring in this paper lie in a finite
dimensional quotient with a single parameter $b$. The Potts representations
are, as before, at the exceptional points of the algebra. The XXZ
representation exists for generic values of the parameters where the parameter
$b$ is related to the twist angle in the spin chain. As in the 2BTL case this
is not a faithful representation. The Potts eigenvalues can again be obtained by combining sectors of
several XXZ chains with different exceptional values of the parameter $b$ \cite{Alcaraz:1988ey,Grimm:1992ni}. We leave a detailed
explanation of the truncation schemes and FSS results to future work.

In the appendix we have given numerical examples which confirm the arguments made in the text.
\section{\TL algebra}
\label{sec:TL}
The \TL algebra \cite{TemperleyLieb,MartinBook}, which we shall denote $TL_N(q)$, is given by generators $e_i$ with $i=1,\cdots,N-1$ obeying the relations:
\bea \label{eqn:TL}
e_i^2&=&(q+q^{-1}) e_i\nonumber \\
e_i e_{i\pm 1} e_i&=&e_i  \\
e_i e_j& =& e_j e_i \quad |i-j|>1 \nonumber 
\eea
Using these relations one finds that there are only a finite number of
distinct words for a given value of $N$. In the next subsections we shall give the
XXZ and Potts representations of the TL algebra.

From the TL algebra a full, spectral parameter dependent, $R$-matrix can be found by
a process of Baxterization \cite{Jones:1990hq}. From the expansion of this transfer matrix one
can obtain, algebraically, the integrable TL Hamiltonian\footnote{As with any integrable model there are
  of course an infinite number of integrable Hamiltonians coming from the
  expansion of the transfer matrix. In this paper we shall only consider the
  simplest one which is linear in the generators}:
\bea \label{eqn:TLintegrable}
H^{TL}=-\sum_{i=1}^{N-1} e_i
\eea
\subsection{Representations}
\subsubsection{XXZ}
On an $L$ site $SU(2)$ spin chain we have a representation of $TL_L(q)$ with arbitrary
parameter $q=e^{i \gamma}$:
\bea \label{eqn:TLXXZ}
e_i= -\half \left\{ \sigma^x_i \sigma^x_{i+1} + \sigma^y_i \sigma^y_{i+1} + \cos \gamma \sigma^z_i \sigma^z_{i+1}  - \cos \gamma + i \sin \gamma \left(\sigma^z_i - \sigma^z_{i+1} \right) \right\}
\eea
%
%
%
\subsubsection{$2$-state Potts (Ising)}
On an $L$ site chain we have a representation of $TL_{2L}(e^{\pi i/4})$ realized by:
\bea
\label{eqn:TLIsing}
e_{2i}&=&\f{1}{\sqrt{2}} \left( 1+ \sigma^z_{i}  \sigma^z_{i+1} \right) \quad
\quad i=1,\cdots,L-1 \nonumber \\
e_{2i-1}&=& \f{1}{\sqrt{2}} \left( 1+ \sigma^x_{i} \right) \quad \quad
i=1,\cdots,L
\eea
Each of these TL generators commutes with the element:
\bea
\label{eqn:Isingsymmetry}
\sigma=\left( \begin{array}{cc} 0 & 1 \\
1 & 0
\end{array} \right) \otimes \left( \begin{array}{cc} 0 & 1 \\
1 & 0
\end{array} \right) \cdots \otimes \left( \begin{array}{cc} 0 & 1 \\
1 & 0
\end{array} \right)
\eea
This obeys $\sigma^2=1$ corresponding to the $Z_2$~($=S_2$) symmetry of the model.
\subsubsection{$3$-state Potts}
On an $L$ site chain we have a representation of $TL_{2L}(e^{\pi i/6})$ realized by:
\begin{eqnarray}
\label{eqn:TLPotts3}
e_{2i}&=&\f{1}{\sqrt{3}} \left( 1+ R_{i}  R^2_{i+1} + R^2_i R_{i+1} \right)
\quad \quad i=1,\cdots,L-1 \nonumber\\
e_{2i-1}&=& \f{1}{\sqrt{3}} \left( 1 + M_i + M^2_i  \right) \quad \quad i=1,\cdots,L
\end{eqnarray}
where the matrices $R,M$ are given by:
\bea
R=\left( \begin{array}{ccc} 1 & 0 &0 \\ 0 & e^{2\pi i/3} & 0 \\ 0 & 0 & e^{4\pi i/3} \end{array} \right) \quad \quad M=\left( \begin{array}{ccc} 0 & 1 &0 \\ 0 & 0 & 1 \\ 1 & 0 & 0 \end{array} \right)
\eea
Each of these TL generators commutes with the elements:
\bea \label{eqn:3Pottssymmetry}
P_1=\left( \begin{array}{ccc} 0 & 1 & 0 \\
1 & 0 & 0 \\
0 & 0 & 1 
\end{array} \right) \otimes \left( \begin{array}{ccc} 0 & 1 & 0 \\
1 & 0 & 0 \\
0 & 0 & 1 
\end{array} \right)  \cdots \otimes \left( \begin{array}{ccc} 0 & 1 & 0 \\
1 & 0 & 0 \\
0 & 0 & 1 
\end{array} \right) \\
P_2=\left( \begin{array}{ccc} 1 & 0 & 0 \\
0 & 0 & 1 \\
0 & 1 & 0 
\end{array} \right) \otimes \left( \begin{array}{ccc} 1 & 0 & 0 \\
0 & 0 & 1 \\
0 & 1 & 0 
\end{array} \right) \cdots \otimes \left( \begin{array}{ccc} 1 & 0 & 0 \\
0 & 0 & 1 \\
0 & 1 & 0 
\end{array} \right) \nonumber
\eea
These obey $P_1^2=1=P_2^2$ and $(P_1 P_2)^3=1$ corresponding to the $D_3$~($=S_3$) symmetry of the model.
\subsubsection{$4$-state Potts}
On an $L$ site chain we have a representation of $TL_{2L}(1)$ realized by:
\begin{eqnarray}
\label{eqn:TLPotts4}
e_{2i}&=&\f{1}{\sqrt{4}} \left( 1+ R_{i}  R^3_{i+1} + R^2_i R^2_{i+1}  + R^3_i
  R_{i+1}\right) \quad \quad i=1,\cdots,L-1 \nonumber\\
e_{2i-1}&=& \f{1}{\sqrt{4}} \left( 1 + M_i + M^2_i +M^3_i \right) \quad \quad i=1,\cdots,L
\end{eqnarray}
where the matrices $R,M$ are given by:
\bea
R=\left( \begin{array}{cccc} 1 & 0 &0&0 \\ 0 & i & 0&0 \\ 0 & 0 & -1 &0\\0 & 0 & 0 &  -i  \end{array} \right) \quad \quad M=\left( \begin{array}{cccc} 0 & 1 &0&0 \\ 0 & 0 & 1&0 \\0&0&0& 1\\  1&0&0 & 0 \end{array} \right)
\eea
These are invariant under a local $S_4$ symmetry generated by:
\bea \label{eqn:4Pottssymmetry}
P_1=\left( \begin{array}{cccc} 
0 & 1 & 0 & 0 \\
1 & 0 & 0 & 0 \\
0 & 0 & 1 & 0 \\
0 & 0 & 0 & 1 \end{array} \right) \nonumber
\otimes
\left( \begin{array}{cccc} 
0 & 1 & 0 & 0 \\
1 & 0 & 0 & 0 \\
0 & 0 & 1 & 0 \\
0 & 0 & 0 & 1 \end{array} \right) 
\cdots \otimes 
\left( \begin{array}{cccc} 
0 & 1 & 0 & 0 \\
1 & 0 & 0 & 0 \\
0 & 0 & 1 & 0 \\
0 & 0 & 0 & 1 \end{array} \right) \\
P_2=\left( \begin{array}{cccc} 
1 & 0 & 0 & 0 \\
0 & 0 & 1 & 0 \\
0 & 1 & 0 & 0 \\
0 & 0 & 0 & 1 \end{array} \right) 
\otimes
\left( \begin{array}{cccc} 
1 & 0 & 0 & 0 \\
0 & 0 & 1 & 0 \\
0 & 1 & 0 & 0 \\
0 & 0 & 0 & 1 \end{array} \right) 
\cdots \otimes 
\left( \begin{array}{cccc} 
1 & 0 & 0 & 0 \\
0 & 0 & 1 & 0 \\
0 & 1 & 0 & 0 \\
0 & 0 & 0 & 1 \end{array} \right) \\
P_3=\left( \begin{array}{cccc} 
1 & 0 & 0 & 0 \\
0 & 1 & 0 & 0 \\
0 & 0 & 0 & 1 \\
0 & 0 & 1 & 0 \end{array} \right) 
\otimes
\left( \begin{array}{cccc} 
1 & 0 & 0 & 0 \\
0 & 1 & 0 & 0 \\
0 & 0 & 0 & 1 \\
0 & 0 & 1 & 0 \end{array} \right)
\cdots \otimes 
\left( \begin{array}{cccc} 
1 & 0 & 0 & 0 \\
0 & 1 & 0 & 0 \\
0 & 0 & 0 & 1 \\
0 & 0 & 1 & 0 \end{array} \right) \nonumber
\eea
On can form TL representations in general $Q$-state Potts models. However for
$Q>4$ these models do not have a second order phase transition and we shall
not discuss them here.
\subsection{Potts spectra within XXZ spectra (review)}
\label{sec:BulkPottsinXXZ}
The integrable TL Hamiltonian (\ref{eqn:TLintegrable}) formed from $TL_{2L}(q)$ is exactly the Hamiltonian for the $L$-site Potts models with free
 boundary conditions where the TL generators are given by (\ref{eqn:TLIsing}),
(\ref{eqn:TLPotts3}), or (\ref{eqn:TLPotts4}) for the Ising, $3$-state and
$4$-state Potts models respectively. It is also the Hamiltonian of the
 $2L$-site integrable XXZ chain with $SU_q(2)$ invariant boundary terms
 where the TL generators are now given by (\ref{eqn:TLXXZ}) with $q=e^{\pi
 i/4}$, $e^{\pi i/6}$ and $1$ for the Ising, $3$-state and
$4$-state Potts models respectively.

We shall use the representation theory of the TL algebra in order to determine which states in the XXZ representation have
energy levels present in the Potts model. For more details on TL representation theory see \cite{MartinBook,Westbury}. In this specific case it is also possible to use the quantum group $SU_q(2)$ which acts as the centralizer of the TL algebra in the XXZ representation \cite{Pasquier:1989kd}. We shall not follow this approach as it cannot be generalized to the boundary and periodic cases. 

Irreducible representations of the TL algebra $TL_N(q)$ are labelled by $V^{(N)}_j$ and are indexed by a `spin' $0 \le j \le \f{N}{2}$
which takes integer values for $N$ even and half-integer values for $N$
odd. These representations can be conveniently encoded in a Bratelli diagram:
\begin{center}
\includegraphics[width=10 cm]{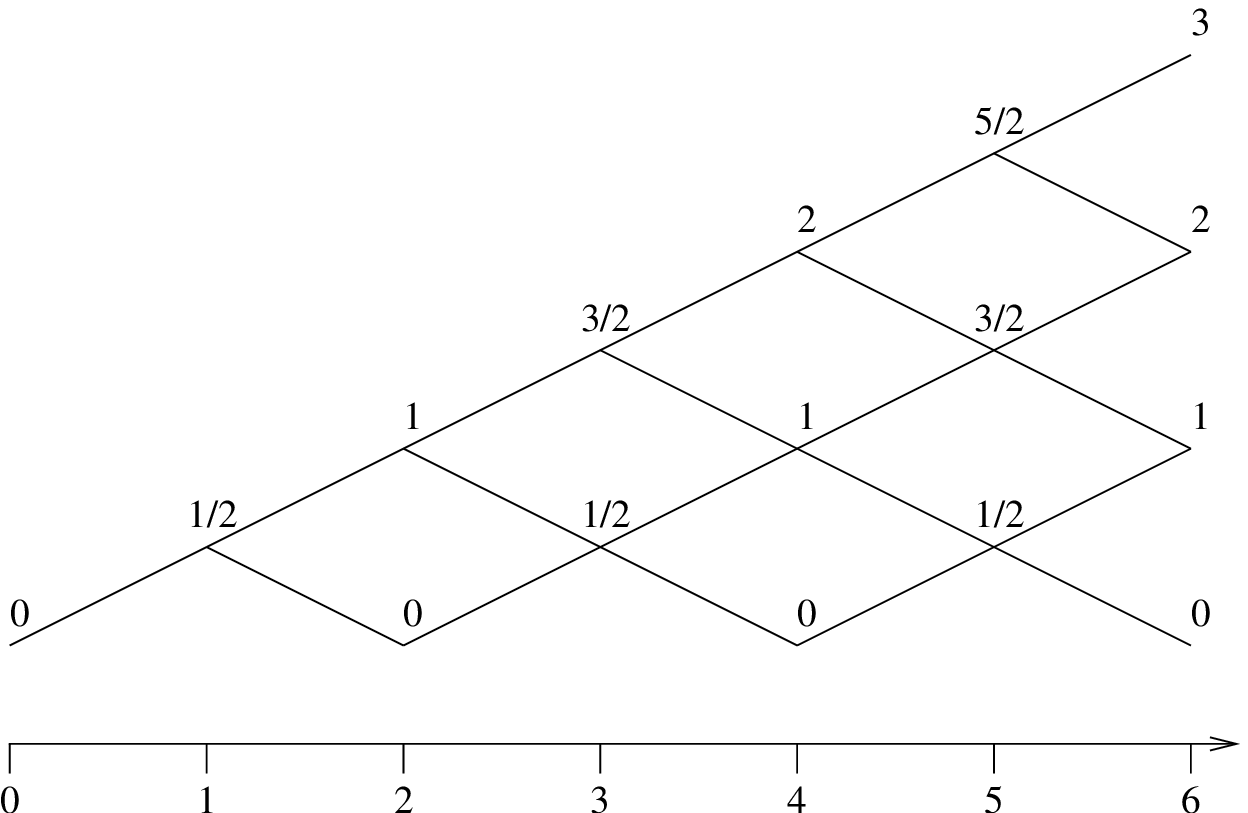}
\label{Bratelli}
\end{center}
On the
horizontal axis the value of $N$ is given. For each value of $N$ the
different irreducible representations $V_j$ that are present are read off. The dimension of the representation $V^{(N)}_j$ is given by
the number of paths that can be drawn between that point and the $0$ at the far
L.H.S. of the diagram. It is given by:
\bea 
\dim V^{(N)}_{j} = \left(\begin{array}{c} N\\ \f{N}{2}-j  \end{array} \right)
- \left(\begin{array}{c} N\\ \f{N}{2}+j+1  \end{array} \right)
\eea
In the $L$-site XXZ representation of $TL_{L}(q)$ each of the irreducible representations $V_j$ occurs with multiplicity $2j+1$. This results in the identity:
\bea \label{eqn:SUq2dimformula}
\sum_{j=0}^{L/2} (2j+1) \dim V^{(L)}_{j}=2^L
\eea
In terms of the quantum group each term in this sum is simply the contribution from the spin $j$ representation of $SU_q(2)$. 

When $q$ is a root of unity we get not just irreducible but also
indecomposable representations. We can now consider a
truncated theory in which we discard the indecomposable representations. For
the case $q^p=\pm 1$ the remaining irreducible representations $V_j$ have $0\le j\le \f{p-2}{2}$.
\begin{itemize}
\item{Ising} 
\label{sec:BulkIsinginXXZ}

For the case of $q=e^{i \pi/4}$ relevant to the Ising model the `good' (irreducible)
representations are $V_0$, $V_{1/2}$, and $V_1$. These can be encoded in a
truncated Bratelli diagram:
\begin{center}
\includegraphics[width=10 cm]{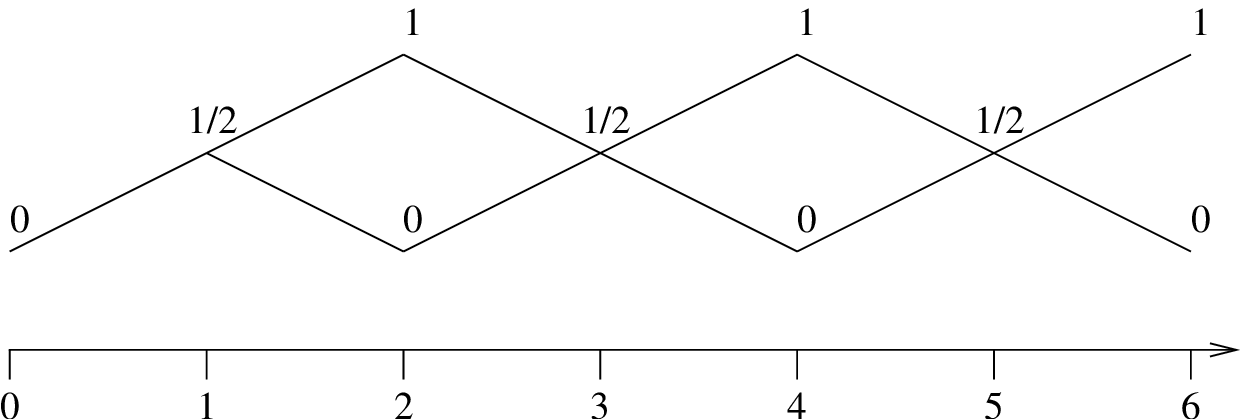}
\end{center}
This is obtained from the full Bratelli diagram on page \pageref{Bratelli} by
simply removing all representations with $j >1$.

The sizes of the irreducible representations can be read off
from the truncated Bratelli diagram (again by counting the paths from the $0$ at the
left hand side) and are given in the table below. In the XXZ chain of length $2L$ the energy levels which also occur in the
Ising model with free boundary conditions come from $V_0$ and $V_1$. One can
see that the total number of states contained in these levels is exactly the number of states
(i.e. $2^L$) present in an $L$-site Ising model.
%
\begin{center}
\begin{tabular}{c|ccc|c}
Length of XXZ chain & \multicolumn{3}{c}{Representation} & Ising Model  \\
    & $V_0$  & $V_{1/2}$  & $V_1$ \\\hline
1 &- &1 &- \\
2 &1 &- &1 & 2\\
3 &- &2 & -\\
4 &2 &- &2 & 4\\
5 &- &4 &- \\
6 &4 &- &4 & 8\\
7 &- &8 &- \\
8 &8 &- &8 & 16 \\
\end{tabular}
\end{center}
%
For instance in the $L=2$ Ising model we have the algebra $TL_4(e^{\pi i/4})$ and a single copy of the $V_1$ and $V_0$ irreducible representations.
\item{$3$-state Potts}
\label{sec:3PottsinXXZ}

For the 3-state Potts model we have 
$q=e^{\pi i/6}$. The `good' representations now have $0 \le j \le 2$ and the truncated Bratelli diagram is given by:
\begin{center}
\includegraphics[width=10 cm]{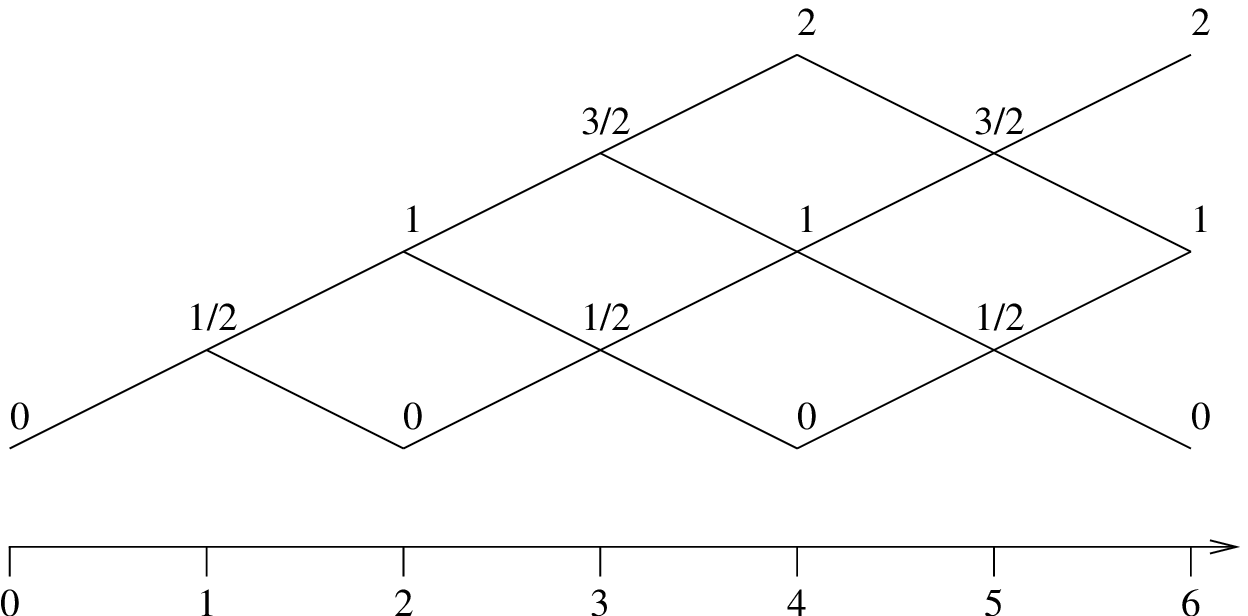}
\end{center}
The $3$-state Potts model has symmetry $S_3$ with irreducible
representations given by singlets and doublets. The
entire spectrum of the Potts model is contained within the XXZ spectrum and
moreover the singlet states from the Potts come from $V_0 \oplus V_2$ and the
doublets from $V_1$. The degeneracies from the Bratelli diagram and the Potts
model are summarized in the following table:
%
%
\begin{center}
\begin{tabular}{c||ccccc|ccc}
Length of & \multicolumn{5}{c}{Representation} &
 \multicolumn{3}{c}{$3$-state Potts Model} \\
 XXZ chain & $V_0$ & $V_{1/2}$  & $V_1$ & $V_{3/2}$ & $V_2$ & Singlets & Doublets & Total Size\\\hline
1 &- &1 &- &- &- \\
2 &1 &- &1 &- &- & 1 & 1 & 3\\
3 &- &2 &- &1 &- \\
4 &2 &- &3 &- &1 & 3 & 3 & 9\\
5 &- &5 &- &4 &- \\
6 &5 &- &9 &- &4 & 9 & 9 & 27 \\
7 &- &14 &- &13 &- \\
8 &14 &- &27 &- &13 &27 & 27 & 81
\end{tabular}
\end{center}
Now in the $L=4$ case the $3$-state Potts representation of $TL_4(e^{\pi i/6})$ contains two copies
of $V_1$ and one each of $V_0$ and $V_2$. 
%
\item{$4$-state Potts}

In the XXZ model this corresponds to $q=1$ (XXX) model. The symmetry
 $SU_q(2)$ becomes simply $SU(2)$ and there is no truncation of the
spectrum. The
spectrum of the a Hamiltonian using the $4$-state Potts representation
(\ref{eqn:TLPotts4}) and
using the XXZ one are found to be identical.
\end{itemize}
In section \ref{se:FSS} we shall discuss the finite size scaling results of
these integrable chains. In the continuum limit they become conformally
invariant and each of the irreducible representations of the TL algebra
becomes an irreducible representation of the Virasoro algebra.
\section{One-boundary \TL algebra}
\renewcommand{\theequation}{\arabic{section}.\arabic{equation}}
\setcounter{equation}{0}  
\label{sec:1BTL}
The one-boundary Temperley-Lieb, or blob, algebra is an interesting generalization of the TL
algebra that has been well studied in the mathematical
literature
\cite{Martin:1992td,Martin:1993jk,MartinWoodcockI,MartinWoodcockII}. In
addition to the TL relations (\ref{eqn:TL}) we have in addition one extra
boundary generator $f_0$ satisfying:
\bea \label{eqn:1BTL}
f_0^2&=&s_1 ~f_0 \nonumber \\
e_1 f_0 e_1&=&e_1  \\
e_i f_0 & =& f_0 e_i \quad i>1 \nonumber
\eea
There is now one extra parameter $s_1$. The space of words is still finite dimensional. Again we can find XXZ and
Potts representations of this algebra.

The integrable Hamiltonian for the 1BTL is given by:
\bea \label{eqn:1BTLintegrable}
H^{1BTL}=-a f_0 -\sum_{i=1}^{N-1} e_i
\eea
where $a$ is an arbitrary coefficient. Therefore by considering this extension
of the TL algebra we are able to describe the addition of an integrable boundary operator.
\subsection{Representations}
\label{sec:1BTLrepns}
\subsubsection{XXZ}
\label{se:1BTLXXZ}
The boundary generator is given by \cite{Martin:1992td}:
\bea
f_0&=&-\half \f{1}{\sin(\omega +\gamma)}\left( -i \cos \omega - \sigma_1^z -
  \cos \phi \sigma_1^x + \sin \phi \sigma_1^y - \sin \omega \right) \nonumber
\eea
This has:
\bea \label{eqn:1BTLs1param}
s_1=\f{\sin \omega}{\sin (\omega+\gamma)}
\eea 
The angle $\phi$ is irrelevant as it can be changed by a rotation of
$\sigma^x_1$ and $\sigma^y_1$ preserving the bulk generators
(\ref{eqn:TLXXZ}). For convenience we shall set $\phi=0$. Then we have:
\bea \label{eqn:XXZe0}
f_0&=&-\half \f{1}{\sin(\omega+\gamma)}\left( -i \cos \omega \sigma_1^z -  \sigma_1^x  - \sin \omega \right) 
\eea
\subsubsection{$2$-state Potts (Ising)}
The most general left boundary generator, which must commute with $e_2$, is
given by $f_0=a+b \sigma^z_1$. However the relation $e_1 f_0 e_1=e_1$ fixes
$a=\f{1}{\sqrt{2}}$. Imposing now $f_0^2=s_1 f_0$ we get, up to action by the
bulk symmetry (\ref{eqn:Isingsymmetry}), a unique solution:
\bea
\label{eqn:1BTLIsing}
f_0=\left( \begin{array}{cc} \sqrt{2} & 0 \\
0 & 0
\end{array} \right) \otimes 1 \otimes \cdots \otimes 1
\eea
This boundary term breaks the $Z_2$ symmetry (\ref{eqn:Isingsymmetry}) of the Ising model. In addition
to $f_0^2=\sqrt{2} f_0$ these two
boundary generators satisfy the additional relation: $f_0 e_1 f_0=f_0$
and so this 1BTL algebra is actually a quotient of the TL algebra $T_{2L+1}(e^{\pi i/4})$. This implies that the spectrum of the one
boundary Hamiltonian (\ref{eqn:1BTLintegrable}) can be extracted from a purely
TL one. We shall return to this point in section \ref{eqn:TruncationTLquotients}.
\subsubsection{$3$-state Potts}
Writing the most general left boundary generator and requiring it to satisfy
the 1BTL algebra gives, up to $S_3$ symmetry (\ref{eqn:3Pottssymmetry}), the following possibilities:
\begin{itemize}
\item{$f_0^2=\sqrt{3} f_0$}
\bea
\label{eqn:1BTL3Pottsfirst}
f_0&=&\left( \begin{array}{ccc} \sqrt{3} & 0 & 0 \\0 & 0 & 0 \\0 & 0 & 0\end{array} \right) \otimes 1 \otimes \cdots \otimes 1 
\eea
This boundary term breaks the $S_3$ symmetry down
to $S_2=Z_2$ as it only invariant under the generator $P_2$ (\ref{eqn:3Pottssymmetry}). We also find, as with the Ising model, that we have $f_0 e_1 f_0=f_0$
and therefore this is again a quotient of the TL algebra $T_{2L+1}(e^{\pi i/6})$.
\item{$f_0^2=\f{\sqrt{3}}{2} f_0$}
\bea
\label{eqn:1BTL3Pottssecond}
f_0&=&\left( \begin{array}{ccc} \f{\sqrt{3}}{2} & 0 & 0 \\0 & \f{\sqrt{3}}{2} & 0 \\0 & 0 & 0\end{array} \right) \otimes 1 \otimes \cdots \otimes 1 
\eea
These boundary conditions again break the $S_3$ symmetry down to
$S_2=Z_2$. This does not satisfy the TL quotient: $f_0 e_1 f_0=f_0$. However it can be related by a linear transformation: $\sqrt{3} -
2 f_0$ to the previous case (\ref{eqn:1BTL3Pottsfirst}). Although this is therefore also therefore also a quotient of $T_{2L+1}(e^{\pi i/6})$ we shall treat them
separately as in the Hamiltonian (\ref{eqn:1BTLintegrable}) this linear transformation reverses the
sign of the boundary term and gives rise to different finite size scaling behaviour -
see section \ref{se:FSS}.
\end{itemize}
\subsubsection{$4$-state Potts}
In this case we have even more possibilities for the left boundary generator:
\begin{itemize}
\item {$f_0^2=\sqrt{4} f_0$}
\bea
\label{eqn:1BTL4Pottsfirst}
f_0&=&\left( \begin{array}{cccc} \sqrt{4} & 0 & 0 & 0 \\0 & 0 & 0 & 0\\0 & 0 & 0 & 0 \\ 0 & 0 & 0 & 0 \end{array} \right) \otimes 1 \otimes \cdots \otimes 1
\eea
This breaks the $S_4$ symmetry down to $S_3$. In this case we find, as with the Ising model, that we have $f_0 e_1 f_0=f_0$
and therefore this is a quotient of the TL algebra $TL_{2L+1}(1)$.
\item{$f_0^2=\f{\sqrt{4}}{2} f_0$}
\bea
f_0&=&\left( \begin{array}{cccc} \f{\sqrt{4}}{2} & 0 & 0 & 0 \\0 & \f{\sqrt{4}}{2}  & 0 & 0\\0 & 0 & 0 & 0 \\ 0 & 0 & 0 & 0 \end{array} \right) \otimes 1 \otimes \cdots \otimes 1 
\eea
These sets of boundary
conditions break the $S_4$ symmetry to $Z_2 \otimes Z_2$. These boundary terms
are not in a TL quotient. 
\item{$f_0^2=\f{\sqrt{4}}{3}$}
\bea
f_0&=&\left( \begin{array}{cccc} \f{\sqrt{4}}{3} & 0 & 0 & 0 \\0 & \f{\sqrt{4}}{3} & 0 & 0\\0 & 0 & \f{\sqrt{4}}{3} & 0 \\ 0 & 0 & 0 & 0 \end{array} \right) \otimes 1 \otimes \cdots \otimes 1
\eea
These sets of boundary conditions break the $S_4$ symmetry to $S_3$. Again, as
in the case of (\ref{eqn:1BTL3Pottssecond}), although these are not in the TL
quotient $f_0 e_1 f_0=f_0$ they can be linearly related to the solutions (\ref{eqn:1BTL4Pottsfirst}).
\end{itemize}
In the $2,3$ and $4$-state Potts models we have given a complete list of possible one-site 
boundary generators of the 1BTL algebra that act only on the first site of
the Potts chain. We shall see in
the next subsection that these all occur at points in which the 1BTL
algebra is non-semisimple and can possess indecomposable representations. These different boundary generators lead to different possible integrable
boundary terms in the Potts Hamiltonian (\ref{eqn:1BTLintegrable}). By directly solving the reflection equation \cite{Cherednik:1985vs,Sklyanin:1988yz}, in the $2,3$ and $4$-state Potts models, we did not find any further \emph{one-site} integrable boundary terms. In \cite{Behrend:2000us} other integrable boundary terms in the Potts models were obtained by considering local boundary terms rather than simply single-site ones. In the continuum conformal field theory description these correspond all the conformally invariant boundary conditions.
\subsection{One-boundary TL representation theory}
\label{se:1BTLrepntheory}
In order to understand the Potts representations we shall use 1BTL
representation theory.

The representation theory of 1BTL has been first worked out by Martin et
al.
\cite{Martin:1992td,Martin:1993jk,MartinWoodcockI,MartinWoodcockII}. For the XXZ representation we have
derived the same truncation schemes \cite{Nichols:2004fb} based on the properties of the centralizer given in \cite{Delius:2001qh,Delius:2002mv,Doikou:2004km}. This approach has a close relation to the
bulk quantum group case and lends support to the fact that the XXZ representation captures all of the irreducible structure of the algebra.

In the study of the 1BTL algebra (\ref{eqn:1BTL}) it is convenient to use the
parameterization for $s_1$ that arises in the XXZ representation:
\bea 
s_1=\f{\sin \omega}{\sin (\omega+\gamma)}
\eea 
In the representation theory a crucial role is played by the relation:
\bea \label{eqn:exceptionalQ}
2 \gamma Q +\omega=\pi {\bf Z}
\eea
There are three different cases depending on the number of solutions of
(\ref{eqn:exceptionalQ}) for $2Q \in {\mathbf Z}$
\begin{itemize}
\item{Generic case:} No solutions.
\item{Critical case:} Only one solution.
\item{Doubly critical case:} Infinitely many solutions.
\end{itemize}
In the generic case the algebra is semi-simple and only possesses irreducible
representations. In the critical or doubly critical cases one also gets indecomposable
representations. In a similar way to the TL case of section
\ref{sec:BulkPottsinXXZ} there is a truncated
sector in which there are only irreducible representations. In the doubly
critical case, relevant for the Potts
models, we always have a finite number of irreducible representations.

In the next three subsections we shall discuss the counting of the number of
`good' states in each of these cases.
\subsubsection{Generic: $\gamma$ and $\omega$ arbitrary}
The irreducible representations of the 1BTL algebra, $W_Q$, are indexed by
$Q=-\f{N}{2}$,$\cdots$, $\f{N}{2}-1$, $\f{N}{2}$. They can be conveniently encoded in
a Bratelli diagram as shown: 
\begin{center}
\includegraphics[width=8 cm]{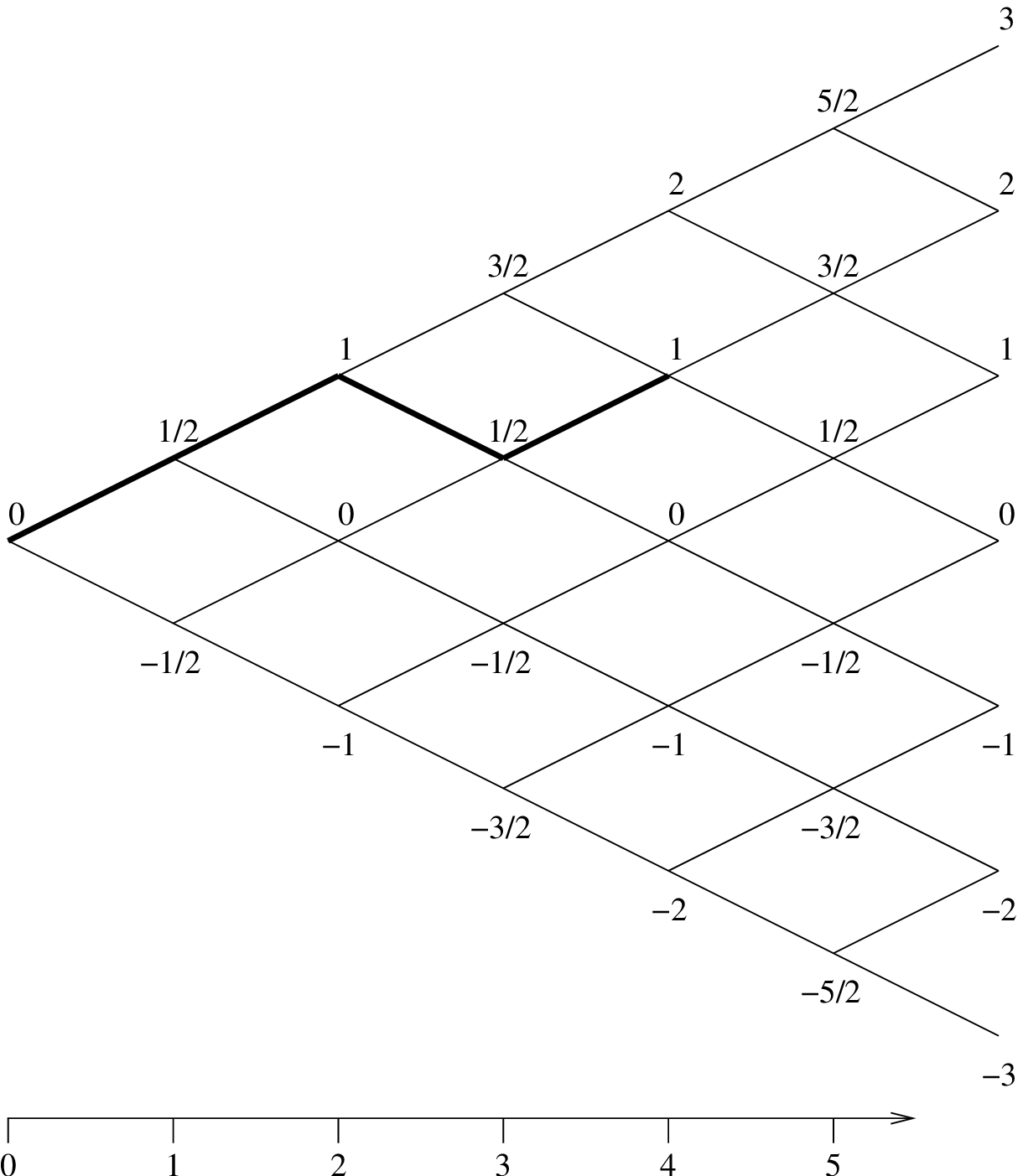}
\end{center}
The system size, $N$, is given on the horizontal axis. One of the paths to the irreducible representation $W_1$ is shown in bold.

The size of irreducible representations $W_Q$ is given by the number of paths from $0$ to that point. In a chain of length $N$ it is given by:
\bea
\left(\begin{array}{c} N\\ \f{N}{2}-Q  \end{array} \right)
\eea
For example in a chain of length $6$ there are $\left(\begin{array}{c} 6\\ 3
  \end{array} \right)=20$ states with $Q=0$. In the $L$-site 1BTL XXZ representation (\ref{eqn:XXZe0}), in contrast to the TL case (\ref{eqn:SUq2dimformula}), each representation $W_Q$ occurs only once:
\bea \label{eqn:1BTLrepnsinXXZ}
\sum_{Q=-L/2}^{L/2} \left(\begin{array}{c} L\\ \f{L}{2}-Q  \end{array} \right) = 2^L
\eea
\subsubsection{Critical: Generic $\gamma$ but $\omega=\gamma {\mathbf Z}$}
If we have:
\bea
\omega=k \gamma
\eea
for some integer $k$ then the 1BTL representation theory becomes
`critical'. Then, as $\gamma$ is generic, the relation (\ref{eqn:exceptionalQ})
has only one solution: $Q=-\f{k}{2}$. Let
us concentrate on the case $k >0 $. 

The space of `good' states becomes truncated from below and the minimum `good'
value of $Q$ is given by $Q=\f{1-k}{2}$. The Bratelli diagram for the case of
$\omega=2 \gamma$ is given by:
\begin{center}
\includegraphics[width=8 cm]{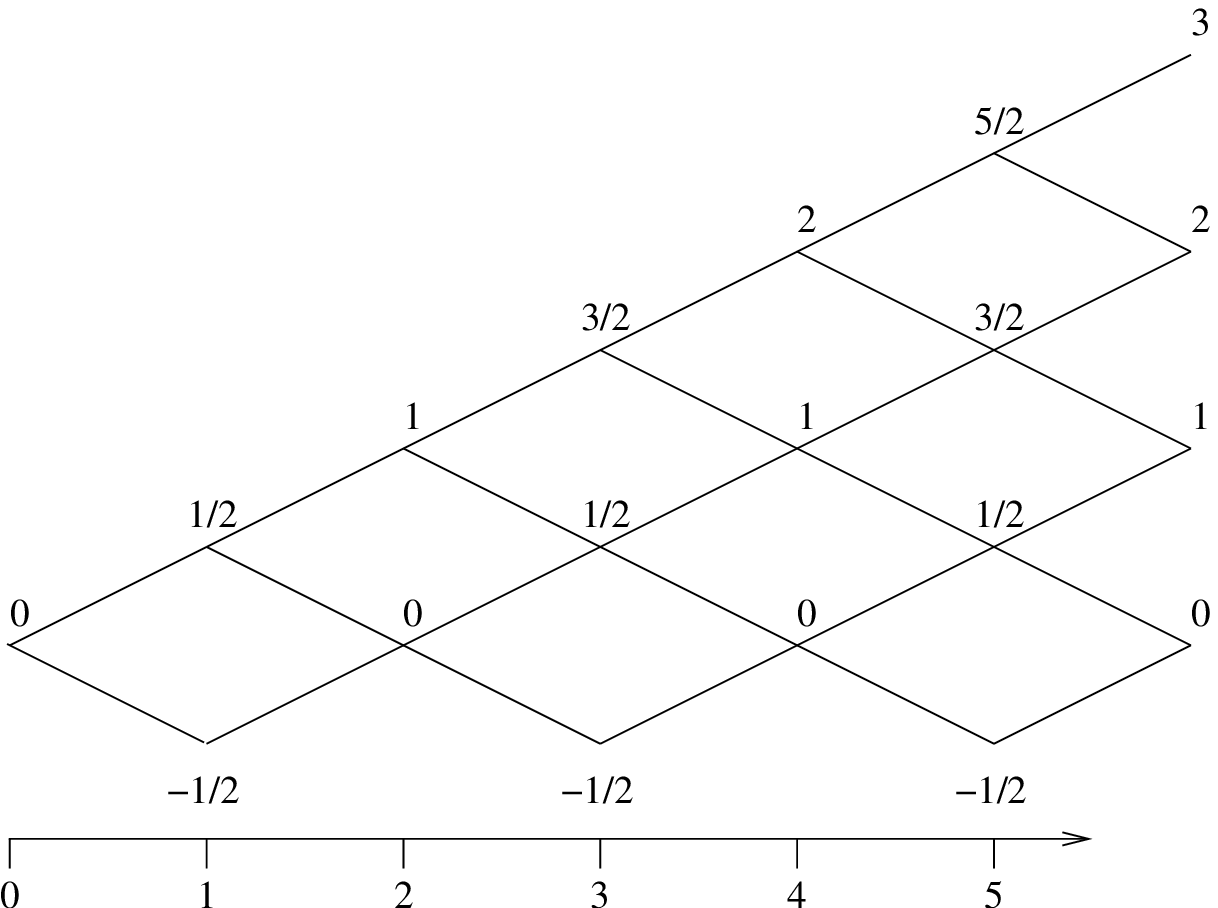}
\end{center}
At a given value of $Q \ge \f{1-k}{2}$ the number of `good' states is given
 by:
\bea \label{eqn:Gammadefined}
\Gamma^{(N)}_{Q} = \left(\begin{array}{c} N\\ \f{N}{2}-Q  \end{array} \right)
- \left(\begin{array}{c} N\\ \f{N}{2}+k+Q  \end{array} \right)
\eea
The first term is the number that would occur in the untruncated diagram and
the second term is the number of paths which go `outside' the
truncated Bratelli diagram. For example in a chain of length $6$ of the $20$
states in the untruncated diagram with $Q=0$ only $20-\left(\begin{array}{c}
    6\\3+2\end{array} \right)=14$ lie inside the truncated diagram shown above.

In the case of $k<0 $ the discussion is very similar but the Bratelli diagram
now becomes truncated from above.
\subsubsection{Doubly critical: $\gamma$ rational multiple of $\pi$ and $\omega=\gamma {\mathbf Z}$}
\label{se:Doublycritical}
This is the most complicated case and also the one that is the most
interesting. Now the relation (\ref{eqn:exceptionalQ}) has an infinite number
of solutions for $2Q \in Z$. Here we shall consider the case:
\bea
\gamma=\f{\pi}{m+1} \quad \quad \omega=k \gamma
\eea
where both $m$ and $k$ are both positive integers. Other rational multiples of
$\pi$ can be dealt with in a similar way.

The lowest `bad' value of $Q$ is given by: $-\f{k}{2}$ and so we
have $Q\ge \f{1-k}{2}$. The highest `bad' value of $Q$ is given by:
$\f{m+1-k}{2}$ and so we have $Q \le \f{m-k}{2}$. Therefore we must have:
\bea
\f{1-k}{2} \le Q \le \f{m-k}{2}
\eea
An example of a doubly truncated Bratelli diagram for the case of
$\gamma=\f{\pi}{6},~\omega=\f{\pi}{3}$ i.e. $m=5,~k=2$ is shown below:
\begin{center}
\includegraphics[width=8 cm]{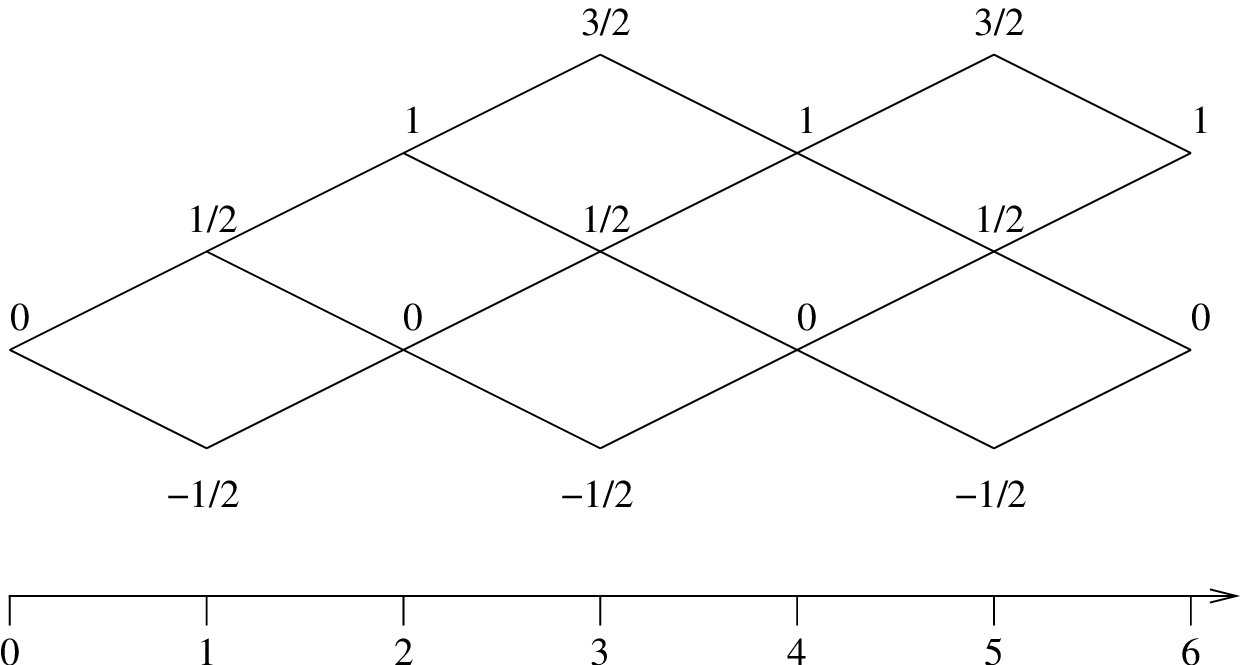}
\end{center}
The calculation of the number of `good' representations, or paths in the
Bratelli diagram, in the doubly critical cases is given by:
\bea \label{eqn:Doublecritcounting}
\Omega^{(N)}_{Q}= \Gamma^{(N)}_{Q} - \Gamma^{(N)}_{m+1-k-Q} + \Gamma^{(N)}_{m+1+Q} - \Gamma^{(N)}_{2(m+1)-k-Q} + \cdots 
\eea
where the $\Gamma^{(N)}_{Q}$ were given before in (\ref{eqn:Gammadefined}). We
shall use this result later when discussing the FSS results. It will
give rise to the infinite number of subtractions which occur in characters of
the minimal conformal field theories. 

In the previous example with $\gamma=\f{\pi}{6},~\omega=\f{\pi}{3}$ the number
of `good' states in the $6$ site case is given by:
\bea
\Omega^{(6)}_{0}&=&\Gamma^{(6)}_{0} - \Gamma^{(6)}_{4-0} + \Gamma^{(6)}_{6+0}
- \Gamma^{(6)}_{10-0} + \cdots \nonumber \\
&=& 14 \\
\Omega^{(6)}_{1}&=&\Gamma^{(6)}_{1} - \Gamma^{(6)}_{4-1} +
\Gamma^{(6)}_{6+1} - \Gamma^{(6)}_{10-1} + \cdots \nonumber \\
&=& 13
\eea
In this case one can easily prove that:
\bea
\Omega^{(2L)}_{0}+\Omega^{(2L)}_{1}=3^L
\eea
As we shall see in the next subsection this is due to the fact that the `good'
states in this model can also be realized in a Potts representation of dimension $3^L$. 
\subsection{Potts spectra within XXZ spectra in the one-boundary cases}
\label{sec:1BTLinPotts}
The integrable 1BTL
Hamiltonian (\ref{eqn:1BTLintegrable}) in the Potts representations
corresponds to Potts models with boundary term added to the left side.
\subsubsection{$2$-state Potts (Ising)}
\label{se:1BTLIsinginXXZ}
There is only one possible boundary term (\ref{eqn:1BTLIsing}). It obeys
$f_0^2=\sqrt{2} f_0$ implying $\gamma=\f{\pi}{4}$ and $\omega=\f{\pi}{2}$. The truncated diagram in this case is given by:
%
\vskip 0.5cm
\begin{center}
\includegraphics[width=8 cm]{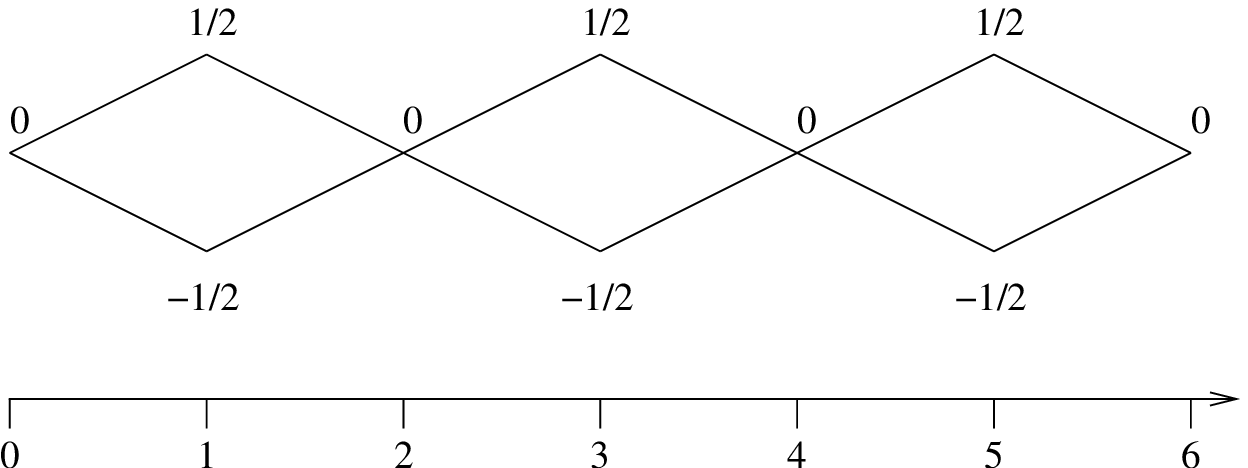}
\end{center}
\vskip 0.5cm
The dimensions of the irreducible representations are given by:
\begin{center}
\begin{tabular}{c||ccc|c}
Length of XXZ chain & \multicolumn{3}{c}{Representation} &
 Ising Model \\
 & $W_{-1/2}$& $W_0$ & $W_{1/2}$ &  Total Size\\\hline
1 & 1 & - & 1 &  \\
2 & - & 2 & - & 2 \\
3 & 2 & - & 2 &  \\
4 & - & 4 & - & 4 \\
5 & 4 & - & 4 &  \\
6 & - & 8 & - & 8 \\
7 & 8 & - & 8 &  \\
8 & - & 16 & - & 16 \\
\end{tabular}
\end{center}
We see that the Ising model contains only the $W_0$ sector. A numerical
example confirming this is given in appendix \ref{sec:1BTLIsing}. 

Note that although
the above table is superficially similar to the TL case of section
\ref{sec:BulkPottsinXXZ} the symmetry group here is different and in the XXZ
model each representation $W_Q$ occurs just once (\ref{eqn:1BTLrepnsinXXZ}). 
\subsubsection{$3$-state Potts}
\label{se:1BTL3PottsinXXZ}
A similar discussion can be made for each of the integrable boundary terms in the $3$-state Potts model. 

The first boundary term (\ref{eqn:1BTL3Pottsfirst}) obeys $f_0^2=\sqrt{3} f_0$
implying $\gamma=\f{\pi}{6}$ and $\omega=\f{2\pi}{3}$. The truncated diagram in this case is given by: 
\vskip 0.5cm
\begin{center}
\includegraphics[width=8 cm]{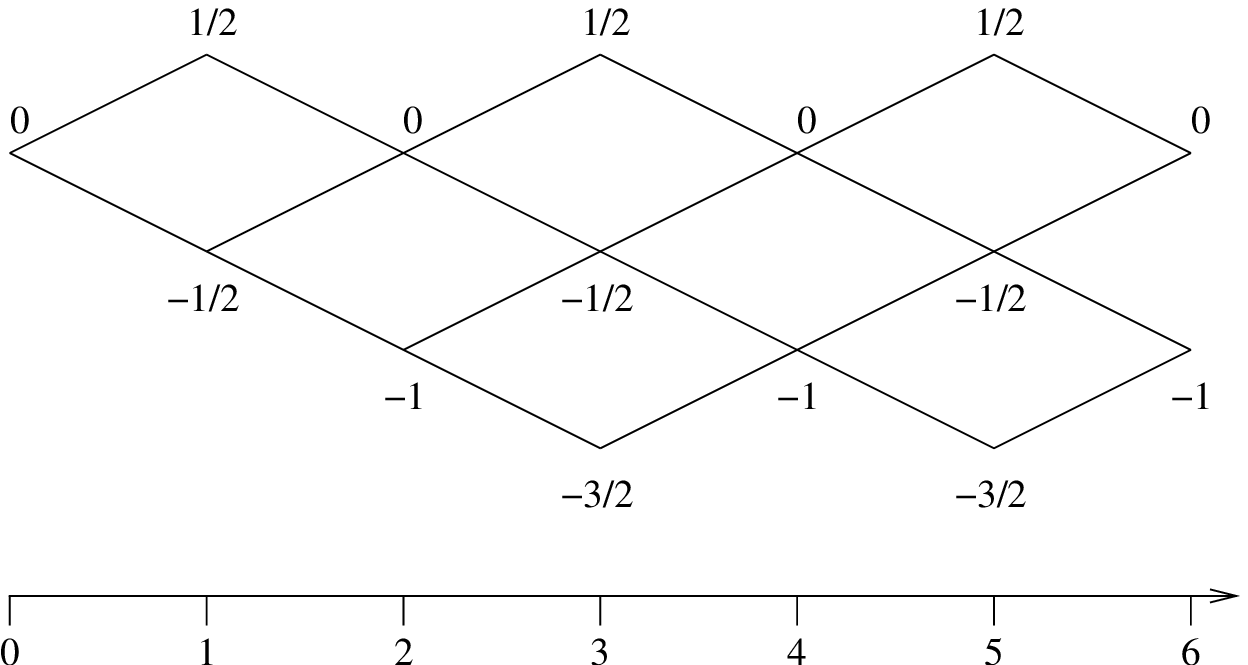}
\vskip 0.5cm
\end{center}
The dimensions of the irreducible representations are given by:
\begin{center}
\begin{tabular}{c||ccccc|c}
Length of XXZ chain & \multicolumn{5}{c}{Representation} &
 Potts Model \\
 & $W_{-3/2} $ & $W_{-1}$ & $W_{-1/2}$ &$W_0$ & $W_{1/2}$ &  Total Size\\\hline
0 &- &- &- &1 &- & 1\\
1 &- &- &1 &- &1 &\\
2 &- &1 &- &2 &- & 3\\
3 &1 &- &3 &- &2 &\\
4 &- &4 &- &5 &- & 9\\
5 &4 &- &9 &- &5 &\\
6 &- &13 &- &14 &- &27\\
7 &13 &- &27 &- &14 &\\
8 &- &40 &- &41 &- &81\\
\end{tabular}
\end{center}
Note that in contrast to the bulk case of section \ref{sec:3PottsinXXZ} all
states are singlets.

The second boundary term (\ref{eqn:1BTL3Pottssecond}) obeys
$f_0^2=\f{\sqrt{3}}{2} f_0$ implying $\gamma=\f{\pi}{6}, \omega=\f{\pi}{3}$
\vskip 0.5cm
\begin{center}
\includegraphics[width=8 cm]{Potts3boundary1.eps}
%
\vskip 0.5cm
\end{center}
Again the dimensions of the irreducible representations can be easily calculated:
\begin{center}
\begin{tabular}{c||ccccc|c}
Length of XXZ chain & \multicolumn{5}{c}{Representation} &
 Potts Model \\
 & $W_{-1/2}$ & $W_{0}$ & $W_{1/2}$ &$W_1$ & $W_{3/2}$ &  Total Size\\\hline
0 &- &1 &- &- &- & 1\\
1 &1 &- &1 &- &- &\\
2 &- &2 &- &1 &- & 3\\
3 &2 &- &3 &- &1 &\\
4 &- &5 &- &4 &- & 9\\
5 &5 &- &9 &- &4 &\\
6 &- &14 &- &13 &- &27\\
7 &14 &- &27 &- &13 &\\
8 &- &41 &- &40 &- &81\\
\end{tabular}
\end{center}
Numerical examples that confirm this picture are given in appendix \ref{sec:1BTL3Potts}.  

In \cite{Nichols:2004fb} an exact spectral equivalence was given between the
one-boundary chain and the general XXZ chain with diagonal boundary
terms. Therefore, for certain values of the parameters, we can find Potts
spectra with a boundary term within the diagonal chain. In section \ref{se:FSS} we shall use this fact
to understand the finite-size scaling of the Potts models with integrable boundary term
added to one end. We shall find that in the finite-size scaling limit
each representation of the 1BTL algebra gives a particular representation of
the Virasoro algebra. For the special cases $\omega=\pm \gamma$ it was shown in
\cite{Nichols:2004fb} that the
boundary generator $f_0$ acts trivially on the `good' truncated sector of
1BTL. This explains why Potts spectra with free boundary conditions could also
be found in the spectrum of particular diagonal chains\cite{Alcaraz:1987uk,Alcaraz:1988ey}.
\section{Two-boundary \TL algebra}
\renewcommand{\theequation}{\arabic{section}.\arabic{equation}}
\setcounter{equation}{0}  
\label{sec:2BTL}
The addition of a second boundary term added at the right hand end proceeds in a similar way to the
1BTL. If we have the 1BTL algebra (\ref{eqn:1BTL}) with $f_0$ and $e_i$ $(i=1,\cdots,N-1)$ then
we introduce an extra element $f_{N}$ satisfying the relations:
\bea
f_{N}^2&=&s_{2} ~f_{N} \nonumber \\
e_{N-1} f_{N} e_{N-1}&=&e_{N-1} \nonumber \\
e_i f_{N} & =& f_{N} e_i \quad 1<i<N-1 \\
f_{0} f_{N} &=& f_{N} f_{0} \nonumber
\eea
The integrable Hamiltonian for the 2BTL is given by:
\bea \label{eqn:2BTLintegrable}
H^{2BTL}=-a f_0 -a' f_{N} -\sum_{i=1}^{N-1} e_i
\eea
where both $a$ and $a'$ are arbitrary coefficients.

In contrast to the TL and 1BTL cases the space of words of the 2BTL is no
longer finite dimensional. However we shall find that the XXZ and Potts
representations lie in particular finite dimensional quotients. Let us first define $I$ and $J$ by:
\begin{itemize}
\item{$N$ even}
\bea 
I&=&e_1 e_3 \cdots e_{N-1} \\
J&=&f_{0}e_2 e_4 \cdots e_{N-2} f_{N}
\eea
\item{$N$ odd}
\bea
I&=&f_{0} e_2 e_4 \cdots e_{N-1} \\
J&=&e_1 e_3 \cdots e_{N-2} f_{N}
\eea
\end{itemize}
In the Potts models the value of $N$ is always even. In this case $I$ and
$J$ are very different as $I$ has no boundary generators in it whereas $J$
contains both. In the odd $N$ cases there is one boundary generator in both $I$ and $J$.

We shall encounter representations which lie in three different types of finite dimensional quotients:
\begin{itemize}
\item{Type I}
%
%
%
\bea \label{eqn:TypeIa}
I J I  =b I
\eea
%
%
%
%
\item{Type II}
\bea \label{eqn:TypeII}
I J I =b I \quad \emph{and} \quad JIJ=bJ
\eea
It is clear by considering $IJIJ$ that we could not have $JIJ=bJ$ and $IJI=b^{'}I$ with $b^{'} \ne b$.
\item{Type III}
%
%
%
%
%
\bea \label{eqn:TypeIIIb}
J=0
\eea
%
%
\end{itemize}
%
It is clear that $II \subset I$ and that $III \subset II_{b=0} \subset
I_{b=0}$.
\subsection{Representations}
\subsubsection{XXZ}
The XXZ representation (on $L$ sites) is given by:
\bea
\label{eqn:XXZfL}
f_0&=&-\half \f{1}{\sin(\omega_1+\gamma)}\left( -i \cos \omega_1 \sigma_1^z -
  \sigma_1^x  - \sin \omega_1 \right) \\
f_L&=&-\half \f{1}{\sin(\omega_2+\gamma)}\left( i \cos \omega_2 \sigma_L^z +
  \cos \phi \sigma_L^x  -\sin \phi \sigma_L^y - \sin \omega_2 \right) \nonumber
\eea
This has:
\bea
s_1=\f{\sin \omega_1}{\sin (\omega_1+\gamma)} \\
s_2=\f{\sin \omega_2}{\sin (\omega_2+\gamma)}
\eea
Note that we have an additional arbitrary angle $\phi$. In the one-boundary
case, discussed in section \ref{se:1BTLXXZ}, in which we have only the generator $f_0$ this angle could be removed by a global $U(1)$ rotation. However in the
two-boundary case we cannot remove them from both ends simultaneously. We find
that the XXZ representation always lives in a Type II quotient and the angle
$\phi$ affects the value of $b$ differently for odd and even values of $L$:
\begin{itemize}
\item{$L$ odd}
\bea \label{eqn:2BTLbLodd}
b=\f{\sin \left(\f{\omega_1-\omega_2-\phi}{2} \right) \sin \left(\f{-\omega_1+\omega_2-\phi}{2}\right)}{\sin(\gamma+\omega_1) \sin(\gamma+\omega_2)}
\eea
\item{$L$ even}
\bea \label{eqn:2BTLbLeven}
b=\f{\sin  \left(\f{\gamma+\omega_1+\omega_2-\phi}{2}\right) \sin  \left(\f{\gamma
  +\omega_1+\omega_2+\phi}{2}\right)}{\sin(\gamma+\omega_1) \sin(\gamma+\omega_2)}
\eea
\end{itemize}
The conventions we use here differ slightly from those of \cite{deGier:2005fx}.
\subsubsection{$2$-state Potts (Ising)}
\label{se:2BTLIsing}
In the $L$ site model if we have two boundary generators $f_0$ and $f_{2L}$ then we have only two
possible cases. These will be labelled $(+,+)$ and $(+,-)$:
\begin{itemize}
\item $(+,+)$: Same boundary terms:
\bea
f_0&=&\f{1}{\sqrt{2}} \left( 1 +  \sigma^z_{1}  \right)
= \left(\begin{array}{cc} \sqrt{2} & 0 \\ 0 & 0 \end{array}\right) \otimes 1 \cdots \otimes 1 \\
f_{2L}&=&\f{1}{\sqrt{2}} \left( 1 +  \sigma^z_{L}  \right)
= 1 \otimes \cdots \otimes 1 \otimes \left(\begin{array}{cc} \sqrt{2} & 0 \\ 0 & 0 \end{array}\right) 
\eea
In this case we find this lies in a Type II quotient: $IJI=\sqrt{2}I, JIJ=\sqrt{2}J$.
\item $(+,-)$: Different boundary terms:
\bea
f_0&=&\f{1}{\sqrt{2}} \left( 1 +  \sigma^z_{1}  \right) 
=\left(\begin{array}{cc} \sqrt{2} & 0 \\ 0 & 0 \end{array}\right) \otimes 1 \cdots \otimes 1\\
f_{2L}&=&\f{1}{\sqrt{2}} \left( 1 -  \sigma^z_{L}  \right)
= 1 \otimes \cdots \otimes 1 \otimes \left(\begin{array}{cc} 0 & 0 \\ 0 & \sqrt{2} \end{array}\right) 
\eea
In this case we find this lies in a Type III quotient: $J=0$. 
\end{itemize}
Other possibilities are equivalent and can be related to these by the global symmetry. In all these cases the bulk $Z_2$ symmetry is completely broken. 
\subsubsection{$3$-state Potts}
\label{se:2BTL3Potts}
We label the three different states of the Potts model by $A,B$ and $C$. The
different possibilities for the boundary terms in the $L$ site model are then labelled according to
the non-zero entries in the matrices: 
\begin{enumerate}
\item{$(A,A)$}
\bea
\label{eqn:3Pottsfirst}
f_0&=&\left( \begin{array}{ccc} \sqrt{3} & 0 & 0 \\0 & 0 & 0 \\0 & 0 & 0\end{array} \right) \otimes 1 \otimes \cdots \otimes 1 \\
f_{2L}&=& 1 \otimes \cdots \otimes 1 \otimes \left( \begin{array}{ccc} \sqrt{3} & 0 & 0 \\0 & 0 & 0 \\0 & 0 & 0\end{array} \right) 
\eea
This breaks $S_3$ to $S_2$. This lies in a Type II quotient: $IJI=\sqrt{3}I,JIJ=\sqrt{3}J$.
\item{$(A,B)$}
\bea
\label{eqn:3PottsAB}
f_0&=&\left( \begin{array}{ccc} \sqrt{3} & 0 & 0 \\0 & 0 & 0 \\0 & 0 & 0\end{array} \right) \otimes 1 \otimes \cdots \otimes 1 \\
f_{2L}&=& 1 \otimes \cdots \otimes 1 \otimes \left( \begin{array}{ccc} 0& 0 & 0 \\0 & \sqrt{3} & 0 \\  0&0&0 \end{array} \right) 
\eea
This breaks $S_3$ completely. This lies in a Type III quotient: $J=0$.
\item{$(A,AB)$}
\bea
f_0&=&\left( \begin{array}{ccc} \sqrt{3} & 0 & 0 \\0 & 0 & 0 \\0 & 0 & 0\end{array} \right) \otimes 1 \otimes \cdots \otimes 1 \\
f_{2L}&=& 1 \otimes \cdots \otimes 1 \otimes \left( \begin{array}{ccc} \f{\sqrt{3}}{2} & 0 & 0 \\0 & \f{\sqrt{3}}{2} & 0\\  0&0&0 \end{array} \right) 
\eea
This breaks $S_3$ completely. This lies in a Type II quotient:
$IJI=\f{\sqrt{3}}{2}I,JIJ=\f{\sqrt{3}}{2}J$.
\item{$(A,BC)$}
\bea
f_0&=&\left( \begin{array}{ccc} \sqrt{3} & 0 & 0 \\0 & 0 & 0 \\0 & 0 & 0\end{array} \right) \otimes 1 \otimes \cdots \otimes 1 \\
f_{2L}&=& 1 \otimes \cdots \otimes 1 \otimes \left( \begin{array}{ccc} 0 & 0 & 0 \\0 & \f{\sqrt{3}}{2} & 0\\  0&0&\f{\sqrt{3}}{2} \end{array} \right) 
\eea
This breaks $S_3$ to $S_2$. This lies in a Type III quotient: $J=0$.
\item{$(AB,AB)$}
\bea
f_0&=&\left( \begin{array}{ccc} \f{\sqrt{3}}{2} & 0 & 0 \\0 & \f{\sqrt{3}}{2} & 0 \\0 & 0 & 0\end{array} \right) \otimes 1 \otimes \cdots \otimes 1 \\
f_{2L}&=& 1 \otimes \cdots \otimes 1 \otimes \left( \begin{array}{ccc} \f{\sqrt{3}}{2} & 0 & 0 \\0 & \f{\sqrt{3}}{2} & 0 \\0 & 0 & 0\end{array} \right)
\eea
This breaks $S_3$ to $S_2$. This lies in a Type I quotient: $IJI=\f{\sqrt{3}}{2}I$.
\item{$(AB,BC)$}
\bea
f_0&=&\left( \begin{array}{ccc} \f{\sqrt{3}}{2} & 0 & 0 \\0 & \f{\sqrt{3}}{2} & 0 \\0 & 0 & 0\end{array} \right) \otimes 1 \otimes \cdots \otimes 1 \\
f_{2L}&=& 1 \otimes \cdots \otimes 1 \otimes \left( \begin{array}{ccc} 0 & 0 & 0 \\0 & \f{\sqrt{3}}{2} & 0\\  0&0&\f{\sqrt{3}}{2} \end{array} \right) 
\eea
This breaks $S_3$ completely. This lies in a Type II quotient: $IJI=\f{\sqrt{3}}{4}I,JIJ=\f{\sqrt{3}}{4}J$.
\end{enumerate}
\subsubsection{$4$-state Potts}
Now we have four states $A,B,C$ and $D$. As in the $3$-state Potts model the
different possibilities for the boundary terms in the $L$ site model are labelled according to
the non-zero entries in the matrices:
\begin{enumerate}
\item{$(A,A)$}
\bea
f_0&=& \left( \begin{array}{cccc} \sqrt{4} & 0 & 0 & 0 \\0 & 0 & 0 & 0\\0 & 0 & 0 & 0 \\ 0 & 0 & 0 & 0 \end{array} \right) \otimes 1 \otimes \cdots \otimes 1 \\
f_{2L}&=& 1 \otimes \cdots \otimes 1 \otimes \left( \begin{array}{cccc} \sqrt{4} & 0 & 0 & 0 \\0 & 0 & 0 & 0\\0 & 0 & 0 & 0 \\ 0 & 0 & 0 & 0 \end{array} \right) 
\eea
This breaks $S_4$ to $S_3$. This lies in a Type II quotient: $IJI=2I,JIJ=2J$.
\item{$(A,B)$}
\bea
f_0&=& \left( \begin{array}{cccc} \sqrt{4} & 0 & 0 & 0 \\0 & 0 & 0 & 0\\0 & 0 & 0 & 0 \\ 0 & 0 & 0 & 0 \end{array} \right) \otimes 1 \otimes \cdots \otimes 1 \\
f_{2L}&=& 1 \otimes \cdots \otimes 1 \otimes \left( \begin{array}{cccc} 0 & 0 & 0 & 0 \\0 & \sqrt{4} & 0 & 0\\0 & 0 & 0 & 0 \\ 0 & 0 & 0 & 0 \end{array} \right) 
\eea
This breaks $S_4$ to $S_2$. This lies in a Type III quotient: $J=0$.
\item{$(A,AB)$}
\bea
f_0&=& \left( \begin{array}{cccc} \sqrt{4} & 0 & 0 & 0 \\0 & 0 & 0 & 0\\0 & 0 & 0 & 0 \\ 0 & 0 & 0 & 0 \end{array} \right) \otimes 1 \otimes \cdots \otimes 1 \\
f_{2L}&=& 1 \otimes \cdots \otimes 1 \otimes \left( \begin{array}{cccc} \f{\sqrt{4}}{2} & 0 & 0 & 0 \\0 & \f{\sqrt{4}}{2} & 0 & 0\\0 & 0 & 0 & 0 \\ 0 & 0 & 0 & 0 \end{array} \right) 
\eea
This breaks $S_4$ to $S_2$. This lies in a Type II quotient: $IJI=I,JIJ=J$.
\item{$(A,BC)$}
\bea
f_0&=& \left( \begin{array}{cccc} \sqrt{4} & 0 & 0 & 0 \\0 & 0 & 0 & 0\\0 & 0 & 0 & 0 \\ 0 & 0 & 0 & 0 \end{array} \right) \otimes 1 \otimes \cdots \otimes 1 \\
f_{2L}&=& 1 \otimes \cdots \otimes 1 \otimes \left( \begin{array}{cccc} 0 & 0 & 0 & 0 \\0 & \f{\sqrt{4}}{2} & 0 & 0\\0 & 0 & \f{\sqrt{4}}{2} & 0 \\ 0 & 0 & 0 & 0 \end{array} \right) 
\eea
This breaks $S_4$ to $S_2$. This lies in a Type III quotient: $J=0$.
\item{$(A,ABC)$}
\bea
f_0&=& \left( \begin{array}{cccc} \sqrt{4} & 0 & 0 & 0 \\0 & 0 & 0 & 0\\0 & 0 & 0 & 0 \\ 0 & 0 & 0 & 0 \end{array} \right) \otimes 1 \otimes \cdots \otimes 1 \\
f_{2L}&=& 1 \otimes \cdots \otimes 1 \otimes \left( \begin{array}{cccc} \f{\sqrt{4}}{3} & 0 & 0 & 0 \\0 & \f{\sqrt{4}}{3}  & 0 & 0\\0 & 0 & \f{\sqrt{4}}{3} & 0 \\ 0 & 0 & 0 & 0 \end{array} \right) 
\eea
This breaks $S_4$ to $S_2$. This lies in a Type II quotient: $IJI=\f{2}{3}I,JIJ=\f{2}{3}J$.
\item{$(A,BCD)$}
\bea
f_0&=& \left( \begin{array}{cccc} \sqrt{4} & 0 & 0 & 0 \\0 & 0 & 0 & 0\\0 & 0 & 0 & 0 \\ 0 & 0 & 0 & 0 \end{array} \right) \otimes 1 \otimes \cdots \otimes 1 \\
f_{2L}&=& 1 \otimes \cdots \otimes 1 \otimes \left( \begin{array}{cccc} 0 & 0 & 0 & 0 \\0 & \f{\sqrt{4}}{3}  & 0 & 0\\0 & 0 & \f{\sqrt{4}}{3} & 0 \\ 0 & 0 & 0 &  \f{\sqrt{4}}{3}\end{array} \right) 
\eea
This breaks $S_4$ to $S_3$. This lies in a Type III quotient: $J=0$.
\item{$(AB,AB)$}
\bea
f_0&=& \left( \begin{array}{cccc} \f{\sqrt{4}}{2} & 0 & 0 & 0 \\0 & \f{\sqrt{4}}{2} & 0 & 0\\0 & 0 & 0 & 0 \\ 0 & 0 & 0 & 0 \end{array} \right) \otimes 1 \otimes \cdots \otimes 1 \\
f_{2L}&=& 1 \otimes \cdots \otimes 1 \otimes \left( \begin{array}{cccc} \f{\sqrt{4}}{2} & 0 & 0 & 0 \\0 & \f{\sqrt{4}}{2} & 0 & 0\\0 & 0 & 0 & 0 \\ 0 & 0 & 0 & 0 \end{array} \right) 
\eea
This breaks $S_4$ to $S_2 \otimes S_2$. This lies in a Type I quotient:
$IJI=I$.
\item{$(AB,BC)$}
\bea
f_0&=& \left( \begin{array}{cccc} \f{\sqrt{4}}{2} & 0 & 0 & 0 \\0 & \f{\sqrt{4}}{2} & 0 & 0\\0 & 0 & 0 & 0 \\ 0 & 0 & 0 & 0 \end{array} \right) \otimes 1 \otimes \cdots \otimes 1 \\
f_{2L}&=& 1 \otimes \cdots \otimes 1 \otimes \left( \begin{array}{cccc} 0 & 0 & 0 & 0 \\0 & \f{\sqrt{4}}{2} & 0 & 0\\0 & 0 & \f{\sqrt{4}}{2} & 0 \\ 0 & 0 & 0 & 0 \end{array} \right) 
\eea
This breaks $S_4$ completely. This lies in a Type II quotient: $IJI=\f{1}{2}I,JIJ=\f{1}{2}J$.
\item{$(AB,CD)$}
\bea
f_0&=& \left( \begin{array}{cccc} \f{\sqrt{4}}{2} & 0 & 0 & 0 \\0 & \f{\sqrt{4}}{2} & 0 & 0\\0 & 0 & 0 & 0 \\ 0 & 0 & 0 & 0 \end{array} \right) \otimes 1 \otimes \cdots \otimes 1 \\
f_{2L}&=& 1 \otimes \cdots \otimes 1 \otimes \left( \begin{array}{cccc} 0 & 0 & 0 & 0 \\0 & 0  & 0 & 0\\0 & 0 & \f{\sqrt{4}}{2} & 0 \\ 0 & 0 & 0 & \f{\sqrt{4}}{2} \end{array} \right) 
\eea
This breaks $S_4$ to $S_2 \otimes S_2$. This lies in a Type III quotient: $J=0$.
\item{$(AB,ABC)$}
\bea
f_0&=& \left( \begin{array}{cccc} \f{\sqrt{4}}{2} & 0 & 0 & 0 \\0 & \f{\sqrt{4}}{2} & 0 & 0\\0 & 0 & 0 & 0 \\ 0 & 0 & 0 & 0 \end{array} \right) \otimes 1 \otimes \cdots \otimes 1 \\
f_{2L}&=& 1 \otimes \cdots \otimes 1 \otimes \left( \begin{array}{cccc} \f{\sqrt{4}}{3} & 0 & 0 & 0 \\0 & \f{\sqrt{4}}{3}  & 0 & 0\\0 & 0 & \f{\sqrt{4}}{3} & 0 \\ 0 & 0 & 0 & 0 \end{array} \right) 
\eea
This breaks $S_4$ to $S_2$. This lies in a Type I quotient: $IJI=\f{2}{3}I$.
\item{$(AB,BCD)$}
\bea
f_0&=& \left( \begin{array}{cccc} \f{\sqrt{4}}{2} & 0 & 0 & 0 \\0 & \f{\sqrt{4}}{2} & 0 & 0\\0 & 0 & 0 & 0 \\ 0 & 0 & 0 & 0 \end{array} \right) \otimes 1 \otimes \cdots \otimes 1 \\
f_{2L}&=& 1 \otimes \cdots \otimes 1 \otimes \left( \begin{array}{cccc} 0 & 0 & 0 & 0 \\0 & \f{\sqrt{4}}{3}  & 0 & 0\\0 & 0 & \f{\sqrt{4}}{3} & 0 \\ 0 & 0 & 0 & \f{\sqrt{4}}{3} \end{array} \right) 
\eea
This breaks $S_4$ to $S_2$. This lies in a Type II quotient: $IJI=\f{1}{3}I,JIJ=\f{1}{3}J$.
\item{$(ABC,ABC)$}
\bea
f_0&=& \left( \begin{array}{cccc} \f{\sqrt{4}}{3} & 0 & 0 & 0 \\0 & \f{\sqrt{4}}{3} & 0 & 0\\0 & 0 & \f{\sqrt{4}}{3} & 0 \\ 0 & 0 & 0 & 0 \end{array} \right) \otimes 1 \otimes \cdots \otimes 1 \\
f_{2L}&=& 1 \otimes \cdots \otimes 1 \otimes \left( \begin{array}{cccc} \f{\sqrt{4}}{3} & 0 & 0 & 0 \\0 & \f{\sqrt{4}}{3} & 0 & 0\\0 & 0 & \f{\sqrt{4}}{3} & 0 \\ 0 & 0 & 0 & 0 \end{array} \right)
\eea
This breaks $S_4$ to $S_3$. This lies in a Type I quotient: $IJI=\f{2}{3}I$.
\item{$(ABC,BCD)$}
\bea
f_0&=& \left( \begin{array}{cccc} \f{\sqrt{4}}{3} & 0 & 0 & 0 \\0 & \f{\sqrt{4}}{3} & 0 & 0\\0 & 0 & \f{\sqrt{4}}{3} & 0 \\ 0 & 0 & 0 & 0 \end{array} \right) \otimes 1 \otimes \cdots \otimes 1 \\
f_{2L}&=& 1 \otimes \cdots \otimes 1 \otimes \left( \begin{array}{cccc} 0 & 0 & 0 & 0 \\0 & \f{\sqrt{4}}{3} & 0 & 0\\0 & 0 & \f{\sqrt{4}}{3} & 0 \\ 0 & 0 & 0 & \f{\sqrt{4}}{3} \end{array} \right)
\eea
This breaks $S_4$ to $S_2$. This lies in a Type I quotient: $IJI=\f{4}{9}I$.
\end{enumerate}
\subsection{Finding Potts spectra within the XXZ spectra}
\label{sec:2BTLPottsfromXXZ}
In the TL and 1BTL algebras the XXZ representation was faithful and therefore
evaluating a Hamiltonian in this representation gives all the
possible eigenvalues allowed by the algebra. Moreover in the Potts models we
understood how to extract the spectra from the XXZ one. In the next section we shall use this fact, together with
 known analytical results from the XXZ model, to deduce finite-size
scaling properties of the Potts models with boundaries.

In the case of the 2BTL the XXZ representation is not a faithful
representation even of the finite dimensional quotients $IJI=bI$ that we
considered\footnote{This can be seen already at two sites where the 
  quotient $IJI=bI$ is $e_1 f_0 f_2 e_1=b e_1$. We have twenty possible words:

${\bf 1}$ , $ f_0$ , $e_1$ , $f_2$ , $f_0 e_1$ , $f_0 f_2$ , $e_1 f_0$ , $ e_1
f_2$ , $f_2 e_1 $ , $f_0 f_2 e_1$ , $e_1 f_0 f_2 $ , $f_0 e_1 f_0$ , $ f_2 e_1
f_0$ , $f_0 e_1 f_2$ , $f_2 e_1  f_2$ , $f_0 e_1 f_0 f_2$ , $f_2 e_1 f_0 f_2$
, $ f_0 f_2 e_1 f_0$ , $ f_0 f_2 e_1 f_2$ ,  $f_0 f_2 e_1 f_0 f_2$.

If we also have the relation $JIJ=bJ$ then we must remove the final word from
this set. The two site XXZ representation is only of dimension $4 \times 4=16$ and so
cannot possibly be faithful.}. Therefore a naive extension of the TL and 1BTL results is
impossible. However, as we shall see, by combining the XXZ results with
several \emph{different} values of the parameter $b$ one can obtain all the
eigenvalues of the Potts models. This phenomena is a consequence of the existence of indecomposable representations that can occur at the exceptional points of the 2BTL algebra.

The first case in which this phenomena occurs is the $3$-state Potts model
($q=e^{\pi i/6}$). Let us consider the two different types of boundary terms given by $s_1=s_2=\sqrt{3}$. In the first (\ref{eqn:3Pottsfirst}) denoted $(A,A)$ an unbroken $Z_2$
symmetry remains. The states can therefore be labelled by their parity under
this symmetry. All the states lie in the quotient $IJI=\sqrt{3}I$. In the
second case $(A,B)$ the
$S_3$ symmetry is fully broken and all the states lie in the quotient $J=0$.

For each of these cases one might hope to be able to obtain these eigenvalues from those of
a single $2L$-site XXZ representation with a particular value of $b$. In appendix \ref{se:Potts2BTLfromXXZ} we have given the eigenvalues of the
Potts chains. We took an inhomogeneous chain to emphasize that integrability
plays no role in our arguments. For the $(A,B)$ case, which has $J=0$, all the eigenvalues are indeed found 
within an XXZ chain with $b=0$. However for the $(A,A)$ case, with $IJI=\sqrt{3} I$,  one can see that only the even parity states (which includes the
ground state) lie in the XXZ model with $b=\sqrt{3}$. We find numerically that
the odd parity states can be obtained by instead taking the XXZ representation
with $b=-\sqrt{3}$. This is not a value of $b$ that occurs in the Potts
model. What is special about these values of $b$ and why should there be
mixing between them?

Recently in \cite{deGier:2005fx} a conjecture was made, based on studies at a
low number of sites, for the critical values of $b$. These are points at which
the 2BTL algebra in the quotient $IJI=bI$ has indecomposable representations. In terms of the
parameterization arising from the XXZ model, (\ref{eqn:2BTLbLodd}) and
(\ref{eqn:2BTLbLeven}), they are given by:
\begin{itemize}
\item{$N$ odd} 
\bea \label{eqn:2BTLexceptionalpointsODD}
\pm \phi= 2 k \gamma + \epsilon_1 w_1 + \epsilon_2 w_2 + 2 \pi Z
\eea
\item{$N$ even} 
\bea \label{eqn:2BTLexceptionalpointsEVEN}
\pm \phi= (2k+1) \gamma + \epsilon_1 w_1 + \epsilon_2 w_2 + 2 \pi Z
\eea
\end{itemize}
where $k$ is a non-negative integer $k < \f{L}{2}$ and 
$\epsilon_1, \epsilon_2 = \pm 1$. Therefore each choice of $k$ gives rise to
four critical points except for $N$ odd and $k=0$ in which case there are only
two. There are precisely $2N$ critical points in total.

The $L$ site Potts model has $N=2L$ even. In the case of the $3$-state Potts
models in which we have boundary parameters $s_1=s_2=\sqrt{3}$
(i.e. $\omega_1=\omega_2=\f{2 \pi}{3}$) we find the only exceptional values are
$b=-\sqrt{3},0,\sqrt{3}$. Firstly we see that, as with the TL and 1BTL
algebras, the Potts representations of 2BTL are at (a subset of) the critical points.

We do not yet have a full understanding of the 2BTL algebra
\cite{WorkInProgress}. Here we shall present arguments based on the structure which we have deduced from studying a low number of sites. At generic
points the full space of 2BTL words in the quotient $IJI=bI$ can be fully
decomposed into irreducible representations:
\bea
{\mathcal V}^{(b)}=V_{XXZ}^{(b)} \oplus V_1 \oplus V_2 \oplus \cdots \oplus V_{2N}
\eea
The XXZ representation, of dimension $2^N$, is generically found to be
irreducible. The quotient also contains $2N$ additional irreducible representations. We have indicated
the dependence of the representations on the parameter $b$. A crucial point is
that all the representations in ${\mathcal V}^{(b)}$ \emph{except} $V_{XXZ}$ are annihilated by $I$ and are therefore insensitive to the value of $b$. Therefore on the full space ${\mathcal V}^{(b)}$ many of the eigenvalues of the 2BTL Hamiltonian (\ref{eqn:2BTLintegrable}) do \emph{not} depend on $b$.

At each of the $2N$ exceptional points, given by
(\ref{eqn:2BTLexceptionalpointsODD}) or (\ref{eqn:2BTLexceptionalpointsEVEN}), we find that a different representation $V_i$ mixes with $V_{XXZ}$ and becomes part of a larger indecomposable representation. This implies that, at an exceptional point, some of the eigenvalues of the 2BTL Hamiltonian on the XXZ space are the same as some of those from another space $V_i$. By allowing $b$ to take different exceptional values we can `read' the eigenvalues from different spaces $V_i$ by just studying the XXZ one. The fact that we had to combine several XXZ representations with different values of $b$ to find all the Potts eigenvalues is simply due to the fact that the Potts representation is composed of several irreducible representations of the 2BTL algebra.

In the tables below we give the sectors of the XXZ model which contribute to
the Potts models with different type of boundary terms.
\begin{itemize}
\item{$2$-Potts: Ising}

In the Ising case the boundary terms are named as in section
\ref{se:2BTLIsing}. In the table below we have indicated with a $*$ the values
of $b$ in the XXZ models in which the
Potts eigenvalues appear. It is important to stress that it is only some of
the XXZ eigenvalues that are used.
\begin{center}
\begin{tabular}{c|cc}
\label{tab:2BTLIsingrepns}
Boundary & \multicolumn{2}{c}{XXZ} \\
Term & $b=0$ & $b=\sqrt{2}$ \\[5pt]
\hline
$(+,+)$ & & *  \\
$(+,-)$ & * & 
\end{tabular}
\end{center}
\item{$3$-state Potts}

In the $3$-state Potts case the boundary terms are named as in section
\ref{se:2BTL3Potts}. In some of the cases the boundary terms are still
invaraint under a $Z_2$ symmetry. In these cases the states can be labelled
by their parity under this symmetry. We have indicated this in the table below
with $\pm$ signs. The ground state is always found to be in the $+$-parity sector.
\begin{center}
\begin{tabular}{l|c|ccccc}
Boundary & Residual & $b=-\sqrt{3}$ & $b=0$ & $b=\f{\sqrt{3}}{4}$ &
$b=\f{\sqrt{3}}{2}$ & $b=\sqrt{3}$ \\
~~~Term &  symmetry & \\
\hline
$(A,A)$ &$Z_2$ & $-$ & & & & $+$ \\
$(A,B)$ & & & * \\
$(A,AB)$ && & & & * & \\
$(A,BC)$ &$Z_2$& &$+$ & & & $-$  \\ 
$(AB,AB)$ &$Z_2$& &$-$ & &$+$ &  \\ 
$(AB,BC)$ && & &* & &
\end{tabular}
\end{center}
\end{itemize}
One can immediately see that in the cases in which there is a residual symmetry
the different parity components are in different irreducible
representations. We shall see later that the number of irreducible
representations of 2BTL algebra present with each boundary term exactly matches with
the number of primary fields in the continuum CFT. This fact was previously
found in the study of particular cases with $\gamma=\f{\pi}{2}$ and
$\f{\pi}{3}$ \cite{deGier:2005fx}.

The Potts representations lie at multi-critical points, generalizing the
notation of doubly critical from section \ref{se:Doublycritical}, in which the
TL and 1BTL subalgebras are also non-semisimple. It remains a challenge to
understand the 2BTL representation theory and the structure of the Potts
representations we have given here. The Bethe Ansatz
for the XXZ model with general boundary terms has (only) been constructed
\cite{ChineseGuys,Nepomechie:2003vv,deGier:2003iu} in the cases in which
they satisfy an additional constraint. This constraint is exactly the same one
defining the critical points of the 2BTL algebra \cite{deGier:2005fx}. An understanding
of the Potts representations will allow one to systematically extract the Potts
spectra, and hence FSS results, from the XXZ one.
\section{Finite size scaling limits}
\renewcommand{\theequation}{\arabic{section}.\arabic{equation}}
\setcounter{equation}{0}  
\label{se:FSS}
\subsection{Temperley-Lieb and 1BTL cases}
In this section we shall discuss the finite size scaling (FSS) limits of the
integrable TL and 1BTL Hamiltonians. We shall be able to use these to discuss FSS limits for the
Potts models with different boundary terms. 
The FSS result for the integrable TL Hamiltonian is well known\cite{Pasquier:1989kd}. The 1BTL case is
new and, as in the TL case, there will be a clear connection between
the representation theory of the finite lattice algebra (i.e. 1BTL) and the
continuum conformal field theory. 

The results for the finite size scaling of the TL chain,
as given in \cite{Pasquier:1989kd}, can be understood in the following
way. We are concerned with the integrable Hamiltonian:
\bea \label{eqn:IntegrableTL}
H=- \f{\gamma}{\pi \sin \gamma} \sum_{i=1}^{N-1} e_i \quad \quad
\eea
The pre-factor is necessary so that the resulting
theory is conformally invariant \cite{Hamer}.

Let us consider the following quantities:
\bea 
\bar{F}_{j;i}(N;\gamma)&=&\f{N}{\pi} \Bigl\{ E_{j;i}(N;\gamma) -
  E_{0;0}(N;\gamma) \Bigr\} \\
{\mathcal F}_{j}(N;\gamma)&=&\sum_i z^{\bar{F}_{j;i}(N;\gamma)}
\eea
where $E_{j;i}(N;\gamma)$ denotes the energy levels, indexed by $i$, of the
Hamiltonian (\ref{eqn:IntegrableTL}) occurring in the irreducible
representation $V_j$. In the case of the XXZ
representation we have $2j+1$ copies of each $V_j$ (\ref{eqn:SUq2dimformula}). The FSS limit is defined as:
\bea \label{eqn:TLFSSdefinition}
{\mathcal F}_{j}(\gamma) = \lim_{N \rightarrow \infty} {\mathcal
  F}_{j}(N;\gamma)
\eea
Now setting:
\bea \label{eqn:gammachoice}
\gamma&=&\f{\pi}{m+1}
\eea
we find the central charge of the theory is given by:
\bea
c=1 - \f{6}{m(m+1)} 
\eea
For $m=3,4,\cdots$ this is the central charge of the $c_{m+1,m}$ minimal
model. In this case there are only a finite number of irreducible
representations with $j \le \f{m-1}{2}$.  These minimal
models have Virasoro degenerate fields \cite{Belavin:1984vu} given by:
\bea \label{eqn:Kacdegeneratefields}
h(r,s)=\f{((m+1)r-ms)^2-1}{4 m(m+1)}
\eea
Then the FSS limit (\ref{eqn:TLFSSdefinition}) of the energy levels of the
integrable Hamiltonian 
(\ref{eqn:IntegrableTL}) occurring in the
irreducible representation $V_j$ are given by $h_{1,2j+1}$ \cite{Pasquier:1989kd}. The main point of
abstracting away from the XXZ chain is that it allows one to
use \emph{any} representation (e.g. Potts) of the TL algebra.

Now we shall discuss the FSS limit for the integrable 1BTL Hamiltonian:
\bea \label{eqn:1BTLHamiltonian}
H^{nd}&=&\f{\gamma}{\pi \sin \gamma} \left(-a f_0-\sum_{i=1}^{N-1} e_i \right)
\eea
Let us begin by using the XXZ representation given in (\ref{eqn:TLXXZ}) and
 (\ref{eqn:XXZe0}). Then parameterizing $a$ by:  
\bea \label{eqn:Definitionofa}
a= \f{2 \sin \gamma \sin(\omega+\gamma)}{\cos \omega + \cos \delta}
\eea
we get the integrable XXZ chain with arbitrary left boundary term:
\bea \label{eqn:Hnd}
H^{nd}&=& \f{\gamma}{\pi \sin \gamma} \left\{ -\f{\sin \gamma}{\cos \omega
    +\cos \delta}\left( i \cos \omega \sigma_1^z + \sigma_1^x  + \sin \omega
  \right)\nonumber \right.\\
&&\!\!\!\!\!\!\!\!\!\left. +\half \left[ \sum_{i=1}^{N-1} \left( \sigma^x_i \sigma^x_{i+1} +
    \sigma^y_i \sigma^y_{i+1} + \cos \gamma \sigma^z_i \sigma^z_{i+1} - \cos
    \gamma \right) + i \sin \gamma \left(\sigma^z_1 - \sigma^z_N \right)
\right] \right\}~~~
\eea
In order to understand the FSS limit of the one-boundary XXZ chain we shall use the spectral equivalence found in
\cite{Nichols:2004fb} between this system and the XXZ chain with purely diagonal boundaries:
\bea \label{eqn:Hd}
H^{d}&=&-\half \f{\gamma}{\pi \sin \gamma}  \left\{ \sum_{i=1}^{N-1}
    \left( \sigma^x_i \sigma^x_{i+1} + \sigma^y_i \sigma^y_{i+1} - \cos \gamma
      \sigma^z_i \sigma^z_{i+1} + \cos \gamma \right) \right.\nonumber\\
&&\left.+\sin \gamma \left[\tan \left(\f{\omega+\delta}{2}\right) \sigma_1^z +
    \tan \left(\f{\omega-\delta}{2} \right)\sigma_N^z  +\f{2 \sin \omega}{\cos
      \omega +\cos \delta} \right] \right\}~~~
\eea
The diagonal Hamiltonian $H^d$ has the obvious local charge:
\bea \label{eqn:Sz}
S^z =\half \sum_{i=1}^N \sigma_i^z
\eea
In \cite{Nichols:2004fb} it was found that for generic values of the parameters the energy levels of
$H^{nd}$ in the sector $W_Q$ (for the definition of 1BTL irreducible sectors $W_Q$ see section \ref{se:1BTLrepntheory}) are exactly the same as the energy levels of $H^{d}$ in the sector with total $S^z$ component equal to $Q$.

The Bethe ansatz for the diagonal chain has been well studied \cite{Alcaraz:1987uk}. The finite size
scaling limit is well understood only if certain numerical truncation schemes are
employed. Here, by using the one-boundary Hamiltonian, we will be able to
understand all of this properly using the 1BTL representation theory of
section \ref{se:1BTLrepntheory}.

The finite size scaling is defined in a similar way to the TL case (\ref{eqn:TLFSSdefinition}). We start
by considering the following quantities:
\bea 
\bar{F}_{Q;i}(N;\gamma,\omega,\delta)&=&\f{N}{\pi} \Bigl\{ E_{Q;i}(N;\gamma,\omega,\delta) -
  E_{0;0}(N;\gamma,\omega=\gamma,\delta) \Bigr\} \\
{\mathcal F}_{Q}(N;\gamma,\omega,\delta)&=&\sum_i z^{\bar{F}_{Q;i}(N;\gamma,\omega,\delta)}
\eea
where again we use $i$ to index the energy levels now within an irreducible representation $W_Q$. The reason for subtracting the ground state energy of the chain with
$\omega=\gamma$ is that this is where the $h_{1,1}=0$ state
lies and we want to keep this energy as our reference state\footnote{It was
  shown in \cite{Nichols:2004fb} that for $\omega=\gamma$ the energy levels of
  the `good' states are the same, up to a constant, as the `good' states
  appearing in the \TL Hamiltonian. Therefore we know the $h_{1,1}=0$ state
  must lie in this sector. }.

The limit of large $N$ is \emph{independent} of $\delta$ as long as $a>0$ (see
\ref{eqn:Definitionofa}) and we get:
\bea \label{eqn:1BTLFSSdefinition}
{\mathcal F}_{Q}(\gamma,\omega)&=& \lim_{N \rightarrow \infty} {\mathcal
  F}_{Q}(N;\gamma,\omega,\delta) \\
&=& z^{\f{(Q- \phi)^2-\alpha^2}{4h}} \prod_{n=1}^{\infty} (1-z^n)^{-1}
\eea
where:
\bea
h&=& \f{1}{4}\left(1- \f{\gamma}{\pi} \right)^{-1} \\
\phi&=&\f{2 h \omega}{\pi} \\
\alpha&=& \f{2 h \gamma}{\pi}
\eea
The value of $\gamma$ is the same as for the bulk theory
(\ref{eqn:gammachoice}). Let us parameterize:
\bea \label{eqn:omega}
\omega &=& r \gamma =  \f{r \pi}{m+1} \\
Q&=&\f{s-r}{2}
\eea
Then we see that:
\bea \label{eqn:FSSgeneric}
{\mathcal F}_{Q}(\gamma,\omega)&=& z^{h(r,s)} \prod_{n=1}^{\infty} (1-z^n)^{-1}
\eea
The leading form of this expression gives the dimension of an operator degenerate at level $rs$ in the Kac-table of the $c_{m+1,m}$ model
(\ref{eqn:Kacdegeneratefields}). However at this stage this is purely formal
as we are at a generic point and so $r,s \notin {\mathbf N}$. In the
exceptional cases we must subtract the `bad' part to obtain the irreducible
characters.

For the `critical' case, where $r,s \in {\mathbf N}$ but $m \notin {\mathbf N}$, we have a single subtraction (\ref{eqn:Gammadefined}):
\bea \label{eqn:FSScritical}
K_{r,s} \equiv {\mathcal F}_{Q}(\gamma,\omega)&=& \left( z^{h(r,s)} - z^{h(r,-s)} \right) \prod_{n=1}^{\infty} (1-z^n)^{-1}
\eea
whereas for the `doubly critical' case there is an infinite number of
subtractions (\ref{eqn:Doublecritcounting}):
\bea \label{eqn:FSSdoublycritical}
\chi_{r,s} &=& K_{r,s} -K_{r,2(m+1)-s} +K_{r,2(m+1)+s}-K_{r,4(m+1)-s} \cdots 
\eea
where $K_{r,s}$ was defined in (\ref{eqn:FSScritical}). This is precisely the character of the $h_{r,s}$ field in the $c_{m+1,m}$
minimal model \cite{Rocha-Caridi}.

We can now rephrase these results in a more algebraic way. We are concerned
with the integrable Hamiltonian (\ref{eqn:1BTLHamiltonian}) with $a>0$. Then the finite size scaling limit, defined in
(\ref{eqn:1BTLFSSdefinition}), of the energy levels occurring in the
irreducible representation $W_Q$ is given by (\ref{eqn:FSSgeneric}),
(\ref{eqn:FSScritical})
 and (\ref{eqn:FSSdoublycritical}) for the generic, critical, and
doubly critical cases (see section \ref{se:1BTLrepntheory}) respectively. As
in the TL case the
advantage of this algebraic viewpoint is that it makes \emph{no} reference to
the actual realization of the 1BTL algebra. If one is presented with a new 1BTL representation, for example
from a loop or RSOS model, once one makes contact with the 1BTL representation
theory one can simply read off the finite size scaling results. 

One should note that the results for the TL case can also be
obtained from the 1BTL algebra in the case $\omega=\pm \gamma$. This is due to
the fact, as shown in \cite{Nichols:2004fb}, that in these cases the boundary operator $f_0$ acts like a
constant on the irreducible 1BTL space and therefore has a trivial effect on
the spectrum. In both cases in the finite size scaling limit we get $h_{1,s}$
fields. This explains why energy levels from the $SU_q(2)$ spectra were also observed in the diagonal chain \cite{Alcaraz:1988vi}.  

Before we discuss the FSS in the Potts models we shall discuss briefly in the
next subsection why the spectra of certain 1BTL and 2BTL problems can be found within the spectra of a TL
Hamiltonian. This will be useful in order to cross-check, and considerably
extend, the TL and 1BTL finite size scaling results.   
\subsection{Truncation in the \TL quotient cases}
\label{eqn:TruncationTLquotients}
We have already seen several examples in the Potts models in which the
one-boundary \TL algebra on $N$ sites is actually a quotient of a TL algebra with one extra generator $T_{N+1}(q)$. In these cases the spectrum of:
\bea \label{eqn:TLquotientchain}
H^{quotient}=-f_0 -\sum_{i=1}^{N-1} e_i
\eea
must be contained in the spectrum of the integrable Hamiltonian for $T_{N+1}(q)$:
\bea
H^{TL}=-\sum_{i=1}^{N} e_i
\eea
when evaluated in a faithful representation like the XXZ one. If the coefficient of the boundary term is different from
that of (\ref{eqn:TLquotientchain}) then we have coincidence between a
boundary chain and an \emph{inhomogeneous} TL chain. In these cases one cannot compare to known
finite size scaling results for integrable TL chains.

A similar situation occurs for certain choices of the the right boundary term,
$f_N$, when the 2BTL algebra also lies in a quotient of TL with two extra
generators. A numerical example illustrating this is given in section \ref{sec:2BTLquotientNumerics}. The existence of these quotients allows one to considerably extend
the FSS results.

In the next two sections we shall use the FSS results and representation theory of TL and 1BTL to derive the finite size scaling of the Potts models with boundary terms. 
\subsection{Ising model}
In this section we shall show how the knowledge of TL and 1BTL representation
theory together with the finite size scaling results for the XXZ model allows
us to derive results in the Ising model. Our conclusions fully agree with the
continuum results of \cite{Cardy:1989ir} if we identify the fixed boundary conditions in the continuum with the appropriate boundary terms added to the Hamiltonian.

The Kac-table for the $c_{4,3}=\f{1}{2}$ minimal model is given below:
\begin{center}
\begin{tabular}{c|cc}
3&$\f{1}{2}$ & $0$
\\[5pt]
2&$\f{1}{16}$ &$\f{1}{16}$
\\[5pt]
1 &$0$ &$\f{1}{2}$
\\[5pt]
\hline
& 1 & 2
\end{tabular}
\end{center}
The entries correspond to the primary fields $h_{r,s}$ as given in
(\ref{eqn:Kacdegeneratefields}). The values of $r$ and $s$ are given on the
horizontal and vertical axis respectively.

As we reviewed in section \ref{sec:BulkIsinginXXZ} the spectra of the $L$-site Ising model
with free boundary conditions is contained in the spectra of the $2L$-site $SU_q(2)$
invariant XXZ model. The `good' states come from $V_0 \oplus V_1$. We know
that the FSS limit of $V_j$ is $h_{1,2j+1}$ which therefore implies that the
finite size scaling limit of the Ising model with free boundary conditions
gives $h_{1,1}=0$ and $h_{1,3}=\f{1}{2}$ fields. This is in agreement with \cite{Cardy:1989ir}.

For the $L$-site Ising model with an integrable boundary term
(\ref{eqn:1BTLIsing}) added to one end (there is only one possibility) we saw in section
\ref{se:1BTLIsinginXXZ} that we have only the
$W_{0}$ representation. Using the FSS scaling results for 1BTL we find this gives $h_{2,2}=\f{1}{16}$. This also
agrees with the results of \cite{Cardy:1989ir}. These states also lie in a quotient of TL
and come from the $V_{1/2}$ representation of the $2L+1$ site XXZ model. We
have denoted this as $(+,{\rm free})^q$ on the table below where the $^q$ superscript indicates the
use of a quotient. Using
the TL finite size scaling results we find this field becomes
$h_{1,2}=\f{1}{16}$. Therefore we have perfect agreement between the two
approaches. In the case of $(+,+)$ boundary terms, defined in section
\ref{sec:2BTLPottsfromXXZ}, these can be understood in
terms of TL quotients, 1BTL quotients, or the 2BTL algebra. In the two
quotient cases we obtain $h=0$. The case of $(+,-)$ boundary terms can be treated in a similar manner
and we obtain $h=\half$. Although we do not yet know the FSS limit of the 2BTL
Hamiltonian (\ref{eqn:2BTLintegrable}) directly we can see that in the Ising
model with $(+,+)$ or $(+,-)$ boundary terms we had just one irreducible
representation of 2BTL (as seen in the table on page \pageref{tab:2BTLIsingrepns}) and one continuum field.

These results are summarised in the following table. We use a $^q$ subscript
to denote the use of a quotient of the algebra:
\begin{table}[h]
\begin{center}
\begin{tabular}{l||c|c|c|c}
 & Length of &  & $L$-site Ising model  & \\
 & XXZ chain & Sector & with boundary term & FSS limit\\
\hline
Temperley-Lieb: & $2L$ ; $2L+2$ & $V_0$ & $({\rm free},{\rm free})$ ; $(+,+)^q$ & $h_{1,1}=0$ \\[5pt]
 & $2L+1$ & $V_{1/2}$ & $(+,{\rm free})^q$ & $h_{1,2}=\f{1}{16}$ \\[5pt]
 & $2L$ ; $2L+2$ & $V_{1}$ & $({\rm free},{\rm free})$ ; $(+,-)^q$  & $h_{1,3}=\f{1}{2}$ \\[5pt]
\hline
1BTL $(f_0^2=\sqrt{2}f_0)$:  & $2L+1$  & $W_{-1/2}$ & $(+,-)^q$ & $h_{2,1}=\f{1}{2}$ \\[5pt]
& $2L$ & $W_0$ & $(+,{\rm free})$ & $h_{2,2}=\f{1}{16}$ \\[5pt]
& $2L+1$ & $W_{1/2}$ & $(+,+)^q$     & $h_{2,3}=0$ \\[5pt]
\end{tabular}
\end{center}
\end{table}
\subsection{$3$-state Potts model}
In this section we shall again use TL and 1BTL representation
theory to derive results in the $3$-state Potts model model. Our conclusions
once again fully agree with those of Cardy \cite{Cardy:1989ir}.

The Kac-Table for $c_{6,5}=\f{4}{5}$ minimal model is given below:
\begin{center}
\begin{tabular}{c|cccc}
5&$3$ &$\f{7}{5}$ & $\f{2}{5}$ & $0$
\\[5pt]
4&$\f{13}{8}$ &$\f{21}{40}$ &$\f{1}{40}$ & $\f{1}{8}$
\\[5pt]
3 &$\f{2}{3}$ &$\f{1}{15}$ &$\f{1}{15}$ & $\f{2}{3}$
\\[5pt]
2 & $\f{1}{8}$ &$\f{1}{40}$ &$\f{21}{40}$ & $\f{13}{8}$
\\[5pt]
1&$0$ &$\f{2}{5}$ &$\f{7}{5}$ &$3$ 
\\[5pt]
\hline
& 1 & 2 & 3 & 4
\end{tabular}
\end{center}
As for the previous Ising example we have labelled $r$ on the horizontal and $s$ on the vertical axis.

For the case of free boundary conditions we saw that the singlets came from
$V_0$ and $V_2$ and the doublets from $V_1$. In the FSS limit this means
singlets come from $h_{1,1}=0$ and $h_{1,5}=3$ and doublets from
$h_{1,3}=\f{2}{3}$.

With the boundary term corresponding to $f_0^2=\sqrt{3} f_0$
(i.e. $\omega=\f{2 \pi}{3}$) we find only the irreducible representations
$W_0$ and $W_{-1}$. In the FSS limit these become $h_{4,4}=\f{1}{8}$ and $h_{4,2}=\f{13}{8}$
respectively. These fields also lie in the $V_{1/2}$ and $V_{3/2}$ irreducible representations of
TL. In the FSS limit these become $h_{1,2}=\f{1}{8}$ and
$h_{1,4}=\f{13}{8}$. Therefore we see total consistency between both
approaches.

For the case of boundary term corresponding to $f_0^2=\f{\sqrt{3}}{2} f_0$
(i.e. $\omega=\f{\pi}{3}$) we find the irreducible representations $W_0$ and $W_1$. In the FSS
limit these become $h_{2,2}=\f{1}{40}$ and $h_{2,4}=\f{21}{40}$. Although
these boundary conditions are linearly related to the previous ones at
$\omega=\f{2 \pi}{3}$ this is of no help as the FSS results are only for the
Hamiltonian (\ref{eqn:IntegrableTL}) and one cannot expect the same result
if some of the terms are reversed.  

The finite size scaling results are shown below where again we use a $^q$ subscript
to denote the use of a quotient of the algebra:
\begin{center}
\begin{tabular}{l||c|c|c|c}
 & Length of &  &  $L$-site $3$-state Potts model & \\
 & XXZ chain & Sector & with boundary term & FSS limit\\
\hline
Temperley-Lieb: & $2L$ ; $2L+2$ & $V_0$ & $({\rm free},{\rm free})$ ; $(A,A)^q$ & $h_{1,1}=0$ \\[5pt]
 & $2L+1$ & $V_{1/2}$ & $(A,{\rm free})^q$  & $h_{1,2}=\f{1}{8}$ \\[5pt]
 & $2L$ & $V_{1}$ & $({\rm free},{\rm free})$ ; $(A,B)^q$  &  $h_{1,3}=\f{2}{3}$ \\[5pt]
 & $2L+1$ & $V_{3/2}$ & $(A,{\rm free})^q$ &  $h_{1,4}=\f{13}{8}$ \\[5pt]
 & $2L$ ; $2L+2$ & $V_{2}$ & $({\rm free},{\rm free})$ ; $(A,A)^q$  &  $h_{1,5}=3$ \\[5pt]
\hline
1BTL $(f_0^2=\sqrt{3}f_0)$: & $2L+1$ & $W_{-3/2}$ &  $(A,A)^q$   & $h_{4,1}=3$ \\[5pt]
& $2L$ & $W_{-1}$ &  $(A,{\rm free})$    & $h_{4,2}=\f{13}{8}$ \\[5pt]
& $2L+1$ & $W_{-1/2}$ &   $(A,B)^q$    & $h_{4,3}=\f{2}{3}$ \\[5pt]
& $2L$ & $W_{0}$ & $(A,{\rm free})$     & $h_{4,4}=\f{1}{8}$ \\[5pt]
& $2L+1$ & $W_{1/2}$ &  $(A,A)^q$    &  $h_{4,5}=0$ \\[5pt]
\hline
1BTL $(f_0^2=\f{\sqrt{3}}{2}f_0)$: & $2L+1$ & $W_{-1/2}$ & $(AB,C)^q$    &  $h_{2,1}=\f{2}{5}$ \\[5pt]
& $2L$ & $W_{0}$ & $(AB,{\rm free})$     & $h_{2,2}=\f{1}{40}$ \\[5pt]
& $2L+1$ & $W_{1/2}$ &  $(AB,A)^q$      & $h_{2,3}=\f{1}{15}$ \\[5pt]
& $2L$ & $W_{1}$ & $(AB,{\rm free})$      & $h_{2,4}=\f{21}{40}$ \\[5pt]
& $2L+1$ & $W_{3/2}$ &  $(AB,C)^q$     & $h_{2,5}=\f{7}{5}$ \\[5pt]
\end{tabular}
\end{center}
These FSS results, some of which can be obtained in several ways, are
consistent with the continuum boundary conformal field results of Cardy\cite{Cardy:1989ir}. We
would like however to stress one important point. We have always referred to
the addition of boundary \emph{terms} rather than boundary conditions. We have
at no point attempted to construct transfer matrices corresponding to fixed
boundary conditions. However consider the Hamiltonian:
\bea
H=-a \left( \begin{array}{ccc} \sqrt{3} & 0 & 0 \\0 & 0 & 0 \\0 & 0 & 0\end{array} \right) -\sum_{i=1}^{N-1} e_i
\eea
with $a>0$. If, under the FSS scaling, $a$ becomes very large then we will
flow towards the situation in which the first spin is fixed into one state.

The other boundary term $AB$ will correspond to a fixed boundary condition
which allows oscillation between $A$ and $B$ \cite{Saleur:1988zx,Cardy:1989ir}.

We have discussed in this section the finite size scaling of the TL and 1BTL integrable Hamiltonians. As we have emphasized throughout this paper the representation theoretic arguments, and subtractions to obtain a unitary theory, are completely algebraic in nature and rely only on the structure of the TL algebra or its appropriate extension. Consider, for example, the Hamiltonian:
\bea
H^{non-critical} = -\sum_{i=1}^{L-1} e_{2i} -\lambda \sum_{i=1}^{L} e_{2i-1}
\eea
For $\lambda \ne 1$ this is the Hamiltonian of the off-critical Potts models. The spectrum of this (non-integrable) theory is now contained within the spectrum of an inhomogeneous XXZ chain and, although the finite size scaling results would be completely different, one uses exactly the same truncation of the lattice TL algebra to obtain the unitary Potts spectrum.
\section{Periodic \TL algebra}
\renewcommand{\theequation}{\arabic{section}.\arabic{equation}}
\setcounter{equation}{0}  
\label{sec:PTL}
This is another extension of the TL algebra, which we shall call periodic
Temperley-Lieb (PTL), in which a new element $e_{N}$ is added to the TL
algebra (\ref{eqn:TL}) satisfying \cite{Levy:1991nc,Martin:1992td,Martin:1993jk}:
\bea
e_{N}^2&=&(q+q^{-1}) e_{N} \nonumber \\
e_i e_{N} e_i&=&e_i  \quad  \quad \quad i=1,N-1 \\
e_{N} e_i e_{N} &=& e_{N}  \quad \quad \quad i=1,N-1 \nonumber \\
e_i e_{N} &=& e_{N} e_i \quad\quad i \ne 1,N-1 \nonumber 
\eea
This algebra, and slight variations of it, have been studied in the mathematical literature \cite{GrahamLehrer,JonesReznikoff}. 

The integrable Hamiltonian for the PTL is given by:
\bea \label{eqn:PTLintegrable}
H^{PTL}=-\sum_{i=1}^{N} e_i
\eea
The space of words is now infinite dimensional. Let us define:
\begin{itemize}
\item{$N$ even}
\bea 
I&=&e_1 e_3 \cdots e_{N-1} \\
J&=& e_2 e_4 \cdots e_{N}
\eea
\item{$N$ odd}
\bea
I&=& e_2 e_4 \cdots e_{N-1} \\
J&=&e_1 e_3 \cdots e_{N-2} e_{N}
\eea
\end{itemize}
In the $L$ site Potts models the value of $N=2L$ is always even. In this paper we shall
encounter representations which lie in three different types of quotients. In terms of the quantities $I$ and $J$ defined here
they take exactly the same form (types I,II and II) as those in the 2BTL case
(\ref{eqn:TypeIa}-\ref{eqn:TypeIIIb}). For $N$ even these quotients are all finite
dimensional however for $N$ odd one must take a further quotient
\cite{Martin:1993jk}.
\subsection{Representations}
\subsubsection{XXZ}
The global symmetry in this case is $U(1)$ and therefore we can consider the toroidal boundary conditions:
\bea
\sigma^{\pm}_{L+1} &=& e^{\pm i \phi} \sigma^{\pm}_1 \\
\sigma^z_{L+1} &=& \sigma^z_1
\eea
where $\sigma^{\pm}=\sigma^x \pm i \sigma^y$. The extra \TL generator is given by:
\bea
\label{eqn:PTLXXZ}
e_L=-\half \left\{ 2 e^{-i \phi} \sigma^+_L \sigma^-_{1} + 2 e^{i \phi}
  \sigma^-_L \sigma^+_{1} + \cos \gamma \sigma^z_L \sigma^z_{1} - \cos \gamma + i \sin
  \gamma  \left(\sigma^z_L - \sigma^z_{1} \right)  \right\}
\eea
Although here we have put all the twist angle $\phi$ dependence into $e_L$ it
is a global property of the system. 

For an even number of sites $L$ we find that this representation lies in a Type II quotient with:
\bea \label{eqn:XXZPTLvalueofb}
b=2+e^{i \phi}+e^{-i \phi}
\eea
For odd number of sites the algebra lies in a Type II quotient only if $e^{i\phi}=1$ with $b=-1$ or $e^{i \phi}=-1$ with $b=1$.

In the XXZ representation all the generators are equivalent and therefore one can introduce an additional `translation' operator $\tau$ with:
\bea
\tau e_i \tau^{-1}=e_{i+1}
\eea
It is clear that this operator commutes with the integrable Hamiltonian
(\ref{eqn:PTLintegrable}) and therefore they can be simultaneously
diagonalized. In terms of the XXZ model the operator $\tau$ is simply the momentum
and it has been used when discussing the spectrum of the model and its finite size scaling \cite{Alcaraz:1988ey}.
\subsubsection{$2$-state Potts (Ising)}
The global symmetry of the bulk model is $S_2=Z_2$. There are two conjugacy classes corresponding to untwisted and twisted periodic boundary conditions:
\begin{itemize}
\item{Untwisted}: $\sigma^z_{L+1}=\sigma^z_1$ 
\bea
e_{2L}= \f{1}{\sqrt{2}} \left( 1+ \sigma^z_{1}  \sigma^z_{L} \right)
\eea
In this case we find this lies in a Type I quotient: $IJI=2I$.
\item{Twisted}: $\sigma^z_{L+1}=-\sigma^z_1$
\bea
e_{2L}= \f{1}{\sqrt{2}} \left( 1- \sigma^z_{1}  \sigma^z_{L} \right)
\eea
In this case we find this lies in a Type III quotient: $J$=0.
\end{itemize}
These conclusions do not depend on the number of sites $L$.
\subsubsection{$3$-state Potts}
The different possible periodic boundary conditions are in correspondence with the conjugacy classes of $S_3$.
\bea
e_{2L}=\f{1}{\sqrt{3}} \left( 1+ R_{L}  R^2_{L+1} + R^2_L R_{L+1} \right)
\eea
\begin{itemize}
\item{123} 
\bea
R_{L+1}=R_1=\left( \begin{array}{ccc} 1 & 0 &0 \\ 0 & e^{2\pi i/3} & 0 \\ 0 & 0 & e^{4\pi i/3} \end{array} \right)
\eea
This lies in a Type I quotient: $IJI=3I$.
\item{(12)3}
\bea
R_{L+1}=\left( \begin{array}{ccc} e^{2\pi i/3} & 0 &0 \\ 0 & 1 & 0 \\ 0 & 0 & e^{4\pi i/3} \end{array} \right)
\eea
This lies in a Type II quotient: $IJI=I$ and $JIJ=J$.
\item{(123)}
\bea
R_{L+1}=\left( \begin{array}{ccc} e^{4\pi i/3}  & 0 &0 \\ 0 & 1 & 0 \\ 0 & 0 & e^{2\pi i/3} \end{array} \right)
\eea
This lies in a Type III quotient: $J=0$.
\end{itemize}
\subsubsection{$4$-state Potts}
The different possible periodic boundary conditions are in correspondence with the conjugacy classes of $S_4$.
\bea
e_{2L}=\f{1}{\sqrt{4}} \left( 1+ R_{L}  R^3_{L+1} + R^2_L R^2_{L+1}  + R^3_L R_{L+1}\right)
\eea
\begin{itemize}
\item{1234}
\bea
R_{L+1}=R_1=\left( \begin{array}{cccc} 1 & 0 &0&0 \\ 0 & i & 0&0 \\ 0 & 0 & -1 &0\\0 & 0 & 0 &  -i  \end{array} \right)
\eea
This lies in a Type I quotient: $IJI=4I$.
\item{(12)34} 
\bea
R_{L+1}=\left( \begin{array}{cccc} i & 0 &0&0 \\ 0 & 1 & 0&0 \\ 0 & 0 & -1 &0\\0 & 0 & 0 &  -i  \end{array} \right)
\eea
This lies in a Type I quotient: $IJI=2I$.
\item{(12)(34)}
\bea
R_{L+1}=R_1=\left( \begin{array}{cccc} i & 0 &0&0 \\ 0 & 1 & 0&0 \\ 0 & 0 & -i &0\\0 & 0 & 0 &  -1  \end{array} \right)
\eea
This lies in a Type III quotient: $J=0$.
\item{(123)4}
\bea
R_{L+1}=R_1=\left( \begin{array}{cccc}-1  & 0 &0&0 \\ 0 & 1 & 0&0 \\ 0 & 0 & i &0\\0 & 0 & 0 &  -i  \end{array} \right)
\eea
This lies in a Type II quotient: $IJI=I,JIJ=J$.
\item{(1234)}
\bea
R_{L+1}=R_1=\left( \begin{array}{cccc}-i  & 0 &0&0 \\ 0 & 1 & 0&0 \\ 0 & 0 & i &0\\0 & 0 & 0 &  -1  \end{array} \right)
\eea
We find  this lies in a Type III quotient: $J=0$.
\end{itemize}
\subsection{Potts within XXZ}
The critical values of $b$ are given in terms of the parameterization arising
in the XXZ representation (\ref{eqn:PTLXXZ}) by \cite{Martin:1992td,Martin:1993jk}:
\bea \label{eqn:PTLcritical}
\phi= 2 \gamma {\bf Z}
\eea
where $q=e^{i \gamma}$. For the Ising model with $q=e^{\pi i/4}$ this gives:
$b=0,2,4$ and for the $3$-state Potts with $q=e^{\pi i/6}$:
$b=0,1,3,4$. We see that all the values realized in the Potts representation
indeed lie at critical points of the PTL algebra.

It has been discussed in
\cite{Alcaraz:1988ey,Grimm:1992ni} how to obtain the spectra of integrable Potts models
numerically from the XXZ ones. For each choice of conjugacy class in the Potts
model one must combine several different twist sectors from the XXZ model. We have used the relation between the twist angle and parameter
$b$ (\ref{eqn:XXZPTLvalueofb}) in the tables below. The
labelling of states $(h,\bar{h})$ is from the continuum CFT. However the
combinations of XXZ sectors, and position of the ground state sector, remain
the same in finite size systems.
\begin{itemize}
\item{$2$-state Potts: Ising}
{\small
\begin{center}
\begin{tabular}{l|l|l}
 & \multicolumn{2}{c}{XXZ} \\
Potts conjugacy class & $b=2$~ $(\phi=\f{\pi}{2})$ &  $b=0$~ $(\phi=\pi)$ \\
\hline
12 ~~~$IJI=2I$& $(0,0)+(\half,\half)$ & $(\f{1}{16},\f{1}{16})$ \\[5pt]
\hline
(12) ~$J=0$ & & $(\f{1}{16},\f{1}{16}) + (0,\half)+ (\half,0)$
\end{tabular}
\end{center}}
\item{$3$-state Potts}
{\small
\begin{center}
\begin{tabular}{l|c|c|c}
 & \multicolumn{3}{c}{XXZ} \\
Potts conjugacy class & $b=3$~$(\phi=\f{\pi}{3})$ & $b=1$~ $(\phi=\f{2\pi}{3})$ & $b=0$~ $(\phi=\pi)$ \\[5pt]
\hline
123 ~~~$IJI=3I$ & $(0,0)+(\f{2}{5},\f{2}{5})$ & & $(0,3)+(3,0)$ \\[5pt]
 & $+(\f{7}{5},\f{7}{5})+(3,3)$ & & $+(\f{2}{5},\f{7}{5})+(\f{7}{5},\f{2}{5})$
 \\[5pt]
 & & & $+(\f{1}{15},\f{1}{15})+(\f{2}{3},\f{2}{3})$ \\[5pt]
\hline
(12)3 ~$IJI=I$ & & $(\f{1}{40},\f{1}{40})+(\f{1}{8},\f{1}{8})$ &
$(\f{1}{40},\f{21}{40})+(\f{21}{40},\f{1}{40})$ \\[5pt]
\quad \quad ~~$JIJ=J$ & & $+(\f{21}{40},\f{21}{40})+(\f{13}{8},\f{13}{8})$ &
$+(\f{1}{8},\f{13}{8})+(\f{13}{8},\f{1}{8})$ \\[5pt]
\hline
(123) ~$J=0$ & & $(0,\f{2}{3})+(\f{2}{5},\f{1}{15})$ &
$(\f{1}{15},\f{1}{15})+(\f{2}{3},\f{2}{3})$ \\[5pt] 
 & & $+(\f{7}{5},\f{1}{15})+(3,\f{2}{3})$ & \\[5pt]
 & & $+(\f{2}{3},0)+(\f{1}{15},\f{2}{5})$ & \\[5pt]
 & & $+(\f{1}{15},\f{7}{5})+(\f{2}{3},3)$ &
\end{tabular}
\end{center}}
\end{itemize}
One can see, in a similar way to the 2BTL case of section \ref{sec:2BTLPottsfromXXZ}, that we have a mixing between
states with different critical values (\ref{eqn:PTLcritical}) of the the parameter $b$. It seems very plausible
that all the mixing, truncations, and FSS results of \cite{Alcaraz:1988ey,Grimm:1992ni} can be deduced
purely from the representation theory of the Periodic \TL algebra but we shall
not attempt this here.
\section{Conclusion}
In this paper we have studied the \TL algebra and its one boundary, two boundary,
and periodic extensions in the Potts and XXZ models. These were used to understand structure and 
relations between the spectra, as well as the finite size scaling behaviour. Throughout this paper, with the exception of the FSS results, we have used algebraic techniques rather than relying on integrability. The same truncation schemes therefore continue to hold in off-critical and inhomogeneous Potts models.

The first, and from our point of view the most basic, algebra that appears is
the \TL algebra (TL) discussed in section \ref{sec:TL}. The Potts
models with free boundary conditions and the XXZ model with $SU_q(2)$ invariant
boundary terms can both be written in terms of the same algebraic \TL
Hamiltonian. The
existence of this underlying algebra allows one to understand relations
between the spectra of these different physical models. Moreover the special
structure of the algebra at exceptional points gives one information on
additional degeneracies as well as providing the possibility of truncation to unitary (e.g. Potts) representations.

We discussed one-boundary, two-boundary, and periodic generalizations
of the TL algebra. In each case, as for \TL, for generic values of the parameters the algebra was semi-simple and
had only irreducible representations. However at exceptional points the
algebra becomes non semi-simple and possessed additional indecomposable
representations.

For the $2,3$ and $4$-state Potts models all the representations of
these algebras were explicitly given. This gives one a simple
classification of possible single-site integrable boundary terms in these models. In every
case one finds that the resulting algebra is at an exceptional
point. In the case of the one-boundary \TL algebra (1BTL) we were able to use
representation theory to understand the Potts representations and extract the
spectra from an XXZ model. For the 2BTL case we found that the spectra could be
obtained from combining the spectra of several different XXZ spin chains. This
fact was again completely algebraic in origin.

In section \ref{se:FSS} we discussed the finite size scaling of the Potts models with
different boundary terms. One can separate the parameters into
those occurring in the algebras and those in the Hamiltonian. This is important
as it is only the parameters of the
algebra which control the structure of the lattice theory as well as its finite
size-scaling limit. In the case of the TL, 1BTL, and every case of the two-boundary
and periodic extensions that we understood, each representation of the
algebra becomes a single representation of the continuum Virasoro
algebra. This connection has been well known for the TL algebra but it is
rather amazing that it continues to hold for all these generalizations. 
\renewcommand\thesection{}
\section{Acknowledgements}
I am grateful for financial support by the E.U. network {\it Integrable models
  and applications: from strings to condensed matter} HPRN-CT-2002-00325. I
  would like to thank V. Rittenberg for many stimulating discussions and to 
  P. Martin for useful correspondence. I would like to thank P. Pearce for interesting discussions and for bringing reference \cite{Behrend:2000us} to my attention.

\renewcommand{\theequation}{\arabic{section}.\arabic{equation}}
\setcounter{equation}{0}  
%
\appendix
%
\renewcommand{\theequation}{\Alph{section}.\arabic{equation}}
\setcounter{equation}{0} 
\section{Numerical results}
%
%
\subsection{One-boundary \TL: Ising}
\label{sec:1BTLIsing}
We take $L=2$, $q=e^{\pi i/4}$ and $w=\f{\pi}{2}$. The Hamiltonian is:
\bea
H=-\alpha_0 f_0 - \sum_{i=1}^{3} \alpha_i e_i
\eea
with:
\bea
(\alpha_0,\cdots,\alpha_3)=(0.68792, 0.72537, 0.33053, 0.95574)
\eea
\begin{center}
\begin{tabular}{l|ccccc|c}
&\multicolumn{5}{c}{$Q$ sector in XXZ} \\
&-2 & -1 & 0 & 1 & 2&  Ising model\\
\hline
&&& -3.32069&  & & -3.32069 \\
&&& &-2.66105 &\\
&&& -2.07891&  & &-2.07891 \\
&&& &-1.92688 &\\
&&-1.87591& & &\\
&&& -1.73886&  & &-1.73886  \\
&&-1.53648& & &\\
&&&-0.97287$^*$ & &-0.97287$^*$\\
&&& &-0.81231 &\\
&&-0.51216& & &\\
&&& -0.49708&  & &-0.49708\\
&&& &-0.36325 &\\
&0$^*$&&0$^*$ & &\\
&&0.106785& & &\\
\hline
\hline
$X(\gamma=\f{\pi}{4})$ & 0 & -1 & 0 & 1 & 0 \\
$X(\gamma=\f{\pi}{4}+\epsilon)$& 4 $\epsilon$ & -1 & 0 & 1 & -4 $\epsilon$
\end{tabular}
\end{center}
The splitting into $Q$-sectors for all the one-boundary examples was performed as follows. On each energy eigenstate the action of the 1BTL centralizer
\cite{Doikou:2004km} was computed. The connection between this centralizer,
$X$, and the 1BTL representation theory has been studied in \cite{Nichols:2004fb}. However, as the eigenvalues of $X$ with $Q=-2,0,2$ are degenerate for
$\gamma=\f{\pi}{4}$, it is difficult to
see from which of these sectors a particular eigenstate of the Hamiltonian came
from. To solve this problem we kept $w=\f{\pi}{2}$ but perturbed  
$\gamma$ slightly away from $\f{\pi}{4}$. As shown in the bottom row of the
table this splits the degeneracy between levels and one can again
clearly see which sectors they came from.

As explained in section \ref{se:1BTLIsinginXXZ} the Ising model contains only the $Q=0$ sector. We
must however discard the doublets which come from indecomposable representations marked with $^*$.
\subsection{One-boundary \TL: $3$-state Potts}
\label{sec:1BTL3Potts}
We take $L=2$, $q=e^{\pi i/6}$ and $w=\f{2 \pi}{3}$. In this case we have
$f_0^2 = \sqrt{3} f_0$. The Hamiltonian is:
\bea
H=-\alpha_0 f_0 - \sum_{i=1}^{3} \alpha_i e_i
\eea
where:
\bea
(\alpha_0,\cdots,\alpha_3)=(0.68792, 0.72537, 0.33053, 0.95574)
\eea
\begin{center}
\begin{tabular}{ccccc|c}
\multicolumn{5}{c}{$Q$ sector in XXZ} \\
-2 & -1 & 0 & 1 & 2&  Potts model\\
\hline
 & & -3.81188 & & & -3.81188 \\
 & & & -3.12797& & \\
 & & -2.49890& & & -2.49890\\
 & & &-2.24237& & \\
 & & -2.12285& &  &-2.12285\\
 & -2.09965& &  & &-2.09965\\
 & -1.80385& &  & &-1.80385\\
 & &-1.19151$^*$ & & -1.19151$^*$ & \\
 & & & -1.05193& & \\
 & -0.71399& &  & &-0.71399\\
 & & -0.70109& &  &-0.70109\\
 & & & -0.63655& & \\
 & & -0.21686& &  &-0.21686\\
 & -0.05829& &  & &-0.05829\\
 0& & & & &
\end{tabular}
\end{center}
We can see that after ignoring any doublets coming from indecomposable
representations we keep only the $Q=-1,0$ sectors to get the Potts
spectrum. This is in exact agreement with the truncated Bratelli diagram in
section \ref{se:1BTL3PottsinXXZ}.

For the second boundary term $\omega=\f{4 \pi}{3}$ we take $L=4$, $q=e^{\pi i/6}$ and $w=\f{4 \pi}{3}$. In this case we have
$f_0^2 = \f{\sqrt{3}}{2} f_0$. The Hamiltonian is:
\bea
H=-\alpha_0 f_0 -\sum_{i=1}^{3} \alpha_i e_i
\eea
where:
\bea
(\alpha_0,\cdots,\alpha_3)=(0.68792, 0.72537, 0.33053, 0.95574)
\eea
\begin{center}
\begin{tabular}{ccccc|c}
\multicolumn{5}{c}{$Q$ sector in XXZ} \\
-2 & -1 & 0 & 1 & 2&  Potts model\\
\hline
& & -3.57991 & & & -3.57991\\
& & & -2.54529 & &  -2.54529\\
& & -2.13969 & & & -2.13969\\
& -2.01982 & & & & \\
& & & -1.87428 & & -1.87428\\
& & -1.75615 & & & -1.75615\\
& -1.68686 & & & & \\
& & -0.85611 & & & -0.85611\\
& & & -0.67184 & & -0.67184\\
& & & & -0.59576 & \\
& -0.42747 & & & & \\
& & -0.42396 & & & -0.42396\\
& & &-0.18014 & & -0.18014\\
& 0.05412 & & & & \\
0$^*$ & & 0$^*$ & & & 
\end{tabular}
\end{center}
We can see that after ignoring any doublets coming from indecomposable
representations we keep only the $Q=0,1$ sectors to get the Potts
spectrum. This is again in exact agreement with the truncated Bratelli diagram
in section \ref{se:1BTL3PottsinXXZ}.
\subsection{Two boundary \TL: $3$-state Potts from XXZ}
\label{se:Potts2BTLfromXXZ}
We take again $L=2$ and the boundary term of the $3$-state Potts model with $f_0^2=\sqrt{3}f_0$ and 
$f_4^2=\sqrt{3} f_4$. Either the two are the same (\ref{eqn:3Pottsfirst}) or they are
different (\ref{eqn:3PottsAB}).
\bea
H=-\alpha_0 f_0 - \alpha_4 f_4 - \sum_{i=1}^{3} \alpha_i e_i 
\eea
with as before:
\bea
(\alpha_0,\cdots,\alpha_4)=(0.68792, 0.72537, 0.33053, 0.95574,0.99239)
\eea
\begin{center}
\begin{tabular}{ccc|cc}
\multicolumn{3}{c}{XXZ representation} & \multicolumn{2}{c}{Potts model} \\
$b=-\sqrt{3}$ & $b=0$ & $b=\sqrt{3}$ & $(A,A)$ & $(A,B)$  \\
\hline
-4.68898 + 0.40346 i    & \underline{-4.71928}   &  \underline{-4.99459} & -4.99459$^+$ & -4.71928 \\
-4.68898 - 0.40346 i    &-4.4806                                                                        & -3.93145 & & \\
-3.64363 + 0.18862 i    & -3.7327    & -3.76919 & & \\
-3.64363 - 0.18862 i    &\underline{-3.47444}       &  \underline{-3.34783} &  -3.34783$^+$ &-3.47444\\
-2.96446                 &               \underline{-2.93234}                                                & -3.10834  & & -2.93234\\
\underline{-2.74265}                                                       &  \underline{-2.84228}                                                     &  -2.91038 & -2.74265$^-$ & -2.84228 \\
-2.21333 - 0.34342 i    & -2.46344 - 0.16479 i  & \underline{-2.74922}  &  -2.74922$^+$ &\\
-2.21333 + 0.34342 i    & -2.46344 + 0.16479 i          & -2.48295 & & \\ 
-2.12550                                                                & \underline{-2.16901}                                                              &  -2.31532 & & -2.16901\\
\underline{-2.01588}                                                        & -1.9856                                                               &  -1.91307  &  -2.01588$^-$ &\\
-1.7012 + 0.07186 i     & -1.71887                                                              &  -1.54222 & & \\
-1.7012 - 0.07186 i     & \underline{-1.38747}                                                              & \underline{-1.46238}  & -1.46238$^+$ & -1.38747\\
-1.24121                                                        &  -1.19151                                                     &       -0.99767  & & \\
\underline{-1.04463}                                                        & \underline{-0.93583}                                                      & -0.81619  & -1.04463$^-$ & -0.93583\\
\underline{-0.591482}                                                       &  \underline{-0.631387}                                                    &  -0.64404 & -0.59148$^-$ & -0.63139 \\
0                                                                                       &  \underline{-0.09190}                                                      & \underline{-0.23527} &  -0.23527$^+$ &-0.09190\\
\end{tabular}
\end{center}
We have underlined the states from the different XXZ representations which are
also to be found in the Potts model. We have labelled the states from $Z_{A,A}$ with their parity under the
residual $Z_2$ symmetry. Note that all the $+$ parity states are found in the
XXZ model with $b=\sqrt{3}$ and the $-$ parity ones in $b=-\sqrt{3}$. All the states
from $Z_{A,B}$ come from the XXZ model with $b=0$.
\subsection{Two boundary \TL: Quotient case}
\label{sec:2BTLquotientNumerics}
We take again $L=2$ and the boundary term of the $3$-state Potts model with $f_0^2=\sqrt{3}f_0$ and 
$f_4^2=\sqrt{3} f_4$. Either the two are the same (\ref{eqn:3Pottsfirst}) or they are
different (\ref{eqn:3PottsAB}).
\bea
H=-\alpha_0 f_0 - \alpha_4 f_4 - \sum_{i=1}^{3} \alpha_i e_i 
\eea
with as before:
\bea
(\alpha_0,\cdots,\alpha_4)=(0.68792, 0.72537, 0.33053, 0.95574,0.99239)
\eea
\begin{center}
\begin{tabular}{cccc|cc}
\multicolumn{4}{c}{$S^z$ sector in XXZ} & \multicolumn{2}{c}{Potts model} \\
$0$ & $\pm 1$ & $\pm 2$ & $\pm 3$ & Same & Different  \\
\hline
-4.99459 & & & & -4.99459$^+$ & \\
-4.71928 & -4.71928  & & & & 4.71928\\
-3.47444 & -3.47444   & & & & -3.47444\\
-3.34783  & & & & -3.34783$^+$ & \\
-2.93234 & -2.93234  & & & & -2.93234\\
-2.84228 & -2.84228  & & & & -2.84228\\
-2.74922  & & & & -2.74922$^+$ & \\ 
-2.74265 & -2.74265 & -2.74265  & & -2.74265$^-$ & \\
-2.16901 & -2.16901  & & & & -2.16901 \\
-2.01588 & -2.01588 &  -2.01588  & & -2.01588$^-$ \\
-1.46238  & & & & -1.46238$^+$ & \\
-1.38747 & -1.38747  & & &  & -1.38747  \\
-1.04463 & -1.04463 & -1.04463  & & -1.04463$^-$ & \\
-0.93583 & -0.93583  & & & & -0.93583 \\
-0.63139 & -0.63139  & & & & -0.63139 \\
-0.59148 & -0.59148 & -0.59148  & & -0.59148$^-$ & \\
-0.23527  & & & & -0.23527$^+$ & \\
-0.09190 & -0.09190  & & & & -0.09190 \\
0,$0^*$ &  0,$0^*$ & 0,$0^*$ & 0 &
\end{tabular}
\end{center}
One can clearly see that the different boundary terms come form $V_1$ and
the same ones from $V_0 \oplus V_2$.
%
%

\end{document}